\documentclass[fleqn, usegraphicx, useAMS, usenatbib]{mn2e}
\usepackage{amsmath, amssymb, aas_macros, subfigure, enumerate}
\begin{document}

\title[Cyclotron lines from accreting neutron stars]{Cyclotron line signatures of thermal and magnetic mountains from accreting neutron stars}
\author[M. Priymak et al.]
{M.~Priymak,$^1$\thanks{E-mail: m.priymak@pgrad.unimelb.edu.au} A.~Melatos,$^1$
\& P.~D.~Lasky$^1$ \\ $^1$School of Physics, University of Melbourne,
Parkville, VIC 3010, Australia}

\maketitle
\begin{abstract}
Cyclotron resonance scattering features (CRSFs) in the X-ray spectrum of an accreting neutron star
are modified differently by accretion mounds sustained by magnetic and thermocompositional gradients.
It is shown that one can discriminate, in principle, between mounds of different physical origins by
studying how the line energy, width, and depth of a CRSF depend on the orientation of the neutron star,
accreted mass, surface temperature distribution, and equation of state.
CRSF signatures including gravitational light bending are computed for both phase-resolved and phase-averaged spectra on the basis of self-consistent
Grad-Shafranov mound equilibria satisfying a global flux-freezing constraint.
The prospects of multimessenger X-ray and gravitational-wave observations with 
future instruments are canvassed briefly.
\end{abstract}

\begin{keywords}
accretion, accretion disks -- gravitational waves -- stars: magnetic field -- stars:
neutron -- pulsars: general
\end{keywords}

\section{Introduction}
\label{sec:introduction}
Recent theoretical modeling has started to elucidate the structural differences between neutron star accretion mounds (`mountains')
sustained by temperature gradients and elastic stresses on the one hand \citep{ushomirsky2000, haskell2006} and magnetic stresses
on the other \citep{brown1998, payne2004, haskell2008, priymak2011, mukherjee2012}.
In a thermal mountain, the temperature gradient at the base of the accreted crust creates kinetic pressure gradients and compositional
(e.g. electron capture) gradients at the surface \citep{ushomirsky2000}.
The structure of the mountain and hence its permanent mass quadrupole moment are limited by the crustal breaking strain \citep{horowitz2009a},
the composition of the crust \citep{haskell2006}, and whether or not the equation of state contains exotic species like 
hyperons or quarks \citep{nayyar2006, haskell2007, johnson2013}.
In a magnetic mountain, in contrast, the mountain structure and mass quadrupole are not limited by the crustal breaking strain.
Magnetic stresses play the dominant role, and the detailed configuration of the surface magnetic field is mainly controlled by the total accreted 
mass, $M_{a}$ \citep{priymak2011}.

In this paper, we investigate how to use X-ray observations of cyclotron lines in accreting neutron stars to distinguish between mountains with different
physical origins. The study is inspired by recent, pioneering calculations of the cyclotron resonance scattering features (CRSFs) produced by
magnetized accretion mounds \citep{bhattacharya2011, bhattacharya2012, mukherjee2012, mukherjee2013a, mukherjee2013b, mukherjee2013c, mukherjee2013d}. 

It is important to find observational signatures to discriminate between thermal and magnetic mountains for two reasons. First, there remains
considerable uncertainty regarding exactly how accreted material spreads away from the accretion column once it hits the stellar surface,
and what this spreading does to the magnetic field. In high-mass X-ray binaries (HMXBs), for example, does a magnetic mountain
form, with material confined at the pole by compressed equatorial magnetic field lines? Or does the accreted material cover the surface
evenly, forming a thermal hot spot at the pole threaded by a dipolar magnetic field?
Even in weakly magnetized systems like low-mass X-ray binaries (LMXBs), there exists evidence from the energies, recurrence times,
and harmonic content of type I thermonuclear bursts that nuclear burning occurs in patches `fenced off' by locally compressed 
magnetic fields much stronger than the global dipole field \citep{bhattacharya2006, payne2006b, misanovic2010, cavecchi2011, chakraborty2012}.
In strongly magnetized systems, like protomagnetars \citep{piro2012, melatos2013} or young pulsars accreting from a fallback disk \citep{wang2006},
the thermomagnetic structure of the accretion mound is completely unknown.

Second, the quadrupole moment of an accretion mound is a gravitational wave source.
In particular, LMXBs are a prime target of ground-based, long-baseline, gravitational-wave detectors like the 
Laser Interferometer Gravitational Wave Observatory (LIGO) and Virgo \citep{riles2013}, 
if they are in a state of torque balance, where the accretion torque equals the gravitational radiation 
reaction torque \citep{bildsten1998, watts2008}\footnote{Other physics may nullify the torque balance scenario, e.g. radiation pressure in the 
inner accretion disk \citep{andersson2005} or episodic cycles in the disk-magnetosphere interaction for a disk
truncated just outside the corotation radius \citep{patruno2012}.}. 
Of course, cyclotron lines have not yet been detected in LMXBs with millisecond periods, the strongest gravitational wave emitters.
Common wisdom holds that this is not surprising because LMXBs are weakly magnetized, and it is true that their inferred
magnetic dipole moments $\mu \sim 10^{26} \ \mathrm{G} \ \mathrm{cm}^{3}$ are low.
However, an alternative scenario is also possible: the cyclotron features are present as a result of the 
`magnetic fences' discussed in the previous paragraph, but they are too weak to be detected by X-ray telescopes
currently in operation. Recent rigourous theoretical calculations have provided support both to the idea of 
locally intense magnetic fields in low-$\mu$ objects \citep{payne2004, payne2006b, vigelius2008a, priymak2011} and the idea
that the resulting cyclotron lines are weak [see \citet{mukherjee2012} and also this paper], adding to the observational
evidence from nuclear burning cited above [see also \citet{patruno2012}]. If so, then interesting multimessenger
(X-ray and gravitational wave) experiments become possible with future observatories.

CRSFs are X-ray absorption lines arising from resonant Compton scattering at quantised Landau energies in the presence 
of a strong magnetic field \citep{meszaros1984, harding1991}. They have been observed in several isolated neutron 
stars \citep{bignami2003, haberl2007} and $17$ accreting X-ray pulsars to date \citep{makishima1999, heindl2004, caballero2012, pottschmidt2012}, 
as well as six unconfirmed candidates \citep{caballero2012, pottschmidt2012}. CRSFs allow an independent and
direct measurement of surface magnetic field strengths between $\sim 10^{12} \ \mathrm{G}$ and $10^{13} \ \mathrm{G}$. 
Several accreting pulsars display multiple, anharmonically-spaced CRSFs \citep{pottschmidt2012}. This anharmonicity is more significant than expected from 
relativistic absorption effects \citep{harding2006} and is attributed instead to local distortions of the magnetic field \citep{nishimura2005}, 
such as those predicted for magnetic mountains.

Numerous results have flowed from studies of CRSFs in accreting pulsars. Some systems such as Her X-1 \citep{staubert2007} and Vela X-1 \citep{fuerst2013} 
show a positive correlation between luminosity and cyclotron line energy, while others such as 4U 0115+63 \citep{mihara2004, nakajima2006} and V 0332+53 \citep{tsygankov2010} show a negative correlation or
none at all [e.g. A 0535+26 \citep{caballero2007}]. These trends are attributed to different accretion rates $\dot{M}$ \citep{basko1976, nelson1993}: 
at high $\dot{M}$, a radiation dominated shock forms in the accretion column, emitting a `fan beam', 
whereas at low $\dot{M}$, a `pencil beam' is emitted from near the surface.
For fast accretors, the luminosity and line energy are anticorrelated because the magnetic field at the shock-front decreases with $\dot{M}$ \citep{burnard1991}. 
CRSF line parameters such as depth, width, and central energy are also observed to vary with pulse phase in 
several sources \citep{kreykenbohm2004, suchy2008, suchy2011}. For example, Her X-1 displays a $\sim 25 \%$ variation in line 
energy and $\sim 100 \%$ variation in line width and depth over a single pulse \citep{klochkov2008}. 
Phase-averaged spectra also exhibit relations between the line energy and the continuum break energy, line width and line energy, and fractional width and line depth \citep{heindl2004}. 

\citet{mukherjee2012} conducted a phase-resolved theoretical study of the effect of accretion-induced deformations of the local magnetic field
on CRSFs, combining the results of Monte-Carlo simulations of photon propagation through magnetically 
threaded plasma \citep{araya1999, araya2000, schonherr2007} with numerical solutions for the magnetohydrostatic structure of the accretion mound \citep{mouschovias1974, brown1998, litwin2001, payne2004, mukherjee2013a}. 
They found that multiple anharmonic CRSFs are consistent with the distorted magnetic field produced by accretion-induced polar burial, with a strong dependence on the viewing geometry. They also found weak (strong) phase dependence 
in the case of symmetric (asymmetric) mounds at the two poles. 
  
In this paper, we apply the above work to the following important practical question:
can one use high-sensitivity X-ray spectra (and possibly simultaneous gravitational-wave data)
to discriminate between a mountain of thermal
and magnetic origin? We find that the answer is a qualified yes; in principle, the two sorts of mountain exhibit different 
instantaneous and phase-averaged CRSF line shapes, and different relations between the line shape and gravitational wave strain,
which may be detectable statistically across the population even if not in individual objects. In the process, we generalize the work of 
Bhattacharya, Mukherjee, and Mignone \citep{mukherjee2012, mukherjee2013a, mukherjee2013b} to include a 
systematic study of CRSF properties as functions of accreted mass, magnetic inclination, viewing angle, 
and equation of state. The paper is structured as follows. 
In Sections \ref{sec:accretion_mound} and \ref{sec:CRSF_model} we describe the numerical models used to generate magnetic mountains and cyclotron line spectra.
We examine the effect of observer orientation and magnetic inclination angles on phase-resolved spectra in Section \ref{sec:orientation}.
We vary the accreted mass and surface temperature distribution in Section \ref{sec:accreted_mass} and the equation of state in Section \ref{sec:equation_of_state}.  
Differences in spectral properties between magnetic and thermo-compositional mountains are discussed in Section \ref{sec:discriminating_mountains}, where we present
a recipe for distinguishing in principle between a magnetic and thermal mountain, partly in preparation for the first searches 
for accreting neutron stars with Advanced LIGO and Virgo.

\section{Accretion mound}
\label{sec:accretion_mound}
We employ the analytical and numerical techniques discussed in detail in \citet{priymak2011}
to compute the equilibrium structure of a magnetized accretion mound, 
by solving the hydromagnetic force balance equation, subject to an integral flux-freezing constraint 
describing the confinement of accreted material to a polar flux tube.
Throughout the paper, we adopt a barotropic equation of state (EOS) [model E in \citet{priymak2011}], unless otherwise stated, 
which parametrises the various pressure 
contributions of a piecewise polytropic EOS into a convenient, $M_{a}$-dependent form and is a fair approximation to the realistic EOS.
The equilibrium calculation is summarized in Section \ref{sec:GS_equilibrium}, 
an example solution is presented in Section \ref{sec:mound_surface}, and the similarities and differences relative to previous work 
are summarized in section \ref{sec:comparison}.

\subsection{Grad-Shafranov equilibrium}
\label{sec:GS_equilibrium}

We define a spherical coordinate system $(r, \theta, \phi)$, where $\theta =
0$ is the magnetic symmetry axis before accretion begins. The magnetic field is given everywhere by
\begin{equation}
\bmath{B} = \frac{\nabla \psi}{r \sin\theta} \times \hat{e}_{\phi},
\label{b_axisymmetric}
\end{equation}
where $\psi(r, \theta)$ is a flux function. In the steady state, the magnetohydrodynamic (MHD)
equations reduce to 
\begin{equation}
\nabla P + \rho \nabla \phi + (\Delta^{2}\psi)\nabla \psi = 0,
\label{mhs_alternative}
\end{equation}
where $\rho(r, \theta)$ is the mass density, $P$ is the pressure, and $\Delta^{2}$ denotes the Grad--Shafranov (GS) operator,
\begin{equation}
\Delta^{2} = \frac{1}{4 \pi r^{2} \sin^{2}\theta} \Bigg[
\frac{\partial^{2}}{\partial r^{2}} + \frac{\sin \theta}{r^{2}}
\frac{\partial}{\partial \theta} \Bigg( \frac{1}{\sin \theta}
\frac{\partial}{\partial \theta} \Bigg) \Bigg].
\label{gs_operator}
\end{equation}

We adopt a general, barotropic equation of state of the form $P(\rho) = K \rho^{\Gamma}$, where $\Gamma$ is the
adiabatic index. For model E in \citet{priymak2011}, both $\Gamma$ and $K$ are scaled with $M_{a}$
to emphasize the dominant pressure contributions acting within the mound. 
We then project equation (\ref{mhs_alternative}) along $\bmath{B}$, and solve for $\psi$
by the method of characteristics. We find 
\begin{equation}
\Delta^{2} \psi = - \frac{\mathrm{d}F(\psi)}{\mathrm{d}\psi} \Bigg\{1 - \frac{(\Gamma - 1) (\phi -
\phi_{0})}{\Gamma K^{1/\Gamma}
[F(\psi)]^{(\Gamma-1)/\Gamma}}\Bigg\}^{1/(\Gamma-1)},
\label{gs_adiabatic}
\end{equation}
where $F(\psi)$ is an arbitrary function of the magnetic flux, and the gravitational potential near the surface takes the form $\phi(r) = G M_{\ast}r/R_{\ast}^{2}$,
with $\phi_{0} = \phi(R_{\ast})$ at the stellar radius $r=R_{\ast}$. 

To establish a one-to-one mapping between the initial (pre-accretion)
and final (post-accretion) states that preserves the flux freezing condition of ideal MHD, 
we require that the final, steady-state, mass-flux distribution $\mathrm{d}M/\mathrm{d}\psi$ equals that of the initial 
state plus the accreted mass. This condition
uniquely determines $F(\psi)$ through
\begin{eqnarray}
\label{F_adiabatic}
F(\psi) & = & \frac{K}{(2 \pi)^{\Gamma}} \Bigg(\frac{\mathrm{d}M}{\mathrm{d} \psi}\Bigg)^{\Gamma} \nonumber \\
& & \times \Bigg[ \int_{C} \mathrm{d}s \ r \sin \theta |\nabla \psi|^{-1} \nonumber \\
& & \times \Bigg\{ 1 - \frac{(\Gamma - 1) (\phi - \phi_{0})}{\Gamma K^{1/\Gamma} [F(\psi)]^{(\Gamma-1)/\Gamma}} \Bigg\}^{1/(\Gamma-1)} \Bigg]^{-\Gamma}.
\end{eqnarray}
The above approach is self-consistent and therefore preferable
to guessing $F(\psi)$ a priori \citep{hameury1983, brown1998, melatos2001, mukherjee2012}. Following earlier work, we
prescribe the mass-flux distribution in one hemisphere to be 
\begin{equation}
M(\psi) = \frac{M_{\mathrm{a}}[1 - \exp(-\psi/\psi_{\mathrm{a}})]}{2[1 - \exp(-b)]},
\label{mass_flux} 
\end{equation}
where $M_{\mathrm{a}}$ is the accreted mass, $\psi_{\ast}$ labels the flux surface
at the magnetic equator, $\psi_{\mathrm{a}}$ labels the field line that closes
just inside the inner edge of the accretion disk, and we write
$b=\psi_{\ast}/\psi_{\mathrm{a}}$. In reality, $M(\psi)$ is determined by the details of the disk-magnetosphere 
interaction \citep{romanova2003, romanova2004}, a topic
outside the scope of this paper.

Equation (\ref{gs_adiabatic}) is solved simultaneously
with equation (\ref{F_adiabatic}) for $\psi(r, \theta)$
using an iterative under-relaxation algorithm combined with a finite-difference
Poisson solver \citep{payne2004, priymak2011}. 
We adopt the following boundary conditions in keeping with previous work:
$\psi(R_{\ast}, \theta) = \psi_{\ast} \sin^{2} \theta$ (surface dipole; magnetic
line tying), $\mathrm{d}\psi/\mathrm{d}r(R_{\mathrm{m}}, \theta) = 0$ (outflow), $\psi(r, 0) = 0$
(straight polar field line), and $\mathrm{d}\psi/\mathrm{d}\theta(r, \pi/2) = 0$ (north-south
symmetry), where $0 \leq r \leq R_{\mathrm{m}}$ and $0 \leq \theta \leq \pi/2$
delimit the computational volume. The outer radius $R_{\mathrm{m}}$ is chosen
large enough to encompass the outer edge of the accreted matter. 
We place the line-tied inner boundary at $R_{\ast}=10^{6} \; \mathrm{cm}$.
The inner boundary is treated as a hard surface for simplicity; subsidence does not change the structure
of the mountain radically \citep{choudhuri2002, wette2010}. Convergence is achieved when the fractional grid-averaged 
$\psi$ residual drops below $10^{-2}$, typically after $10^{3}$ iterations on a $256 \times 256$ grid \citep{payne2004}.

\subsection{Mound surface}
\label{sec:mound_surface}

\begin{figure*}
\begin{minipage}{170mm} 
\subfigure
{
	\includegraphics[width=84mm]{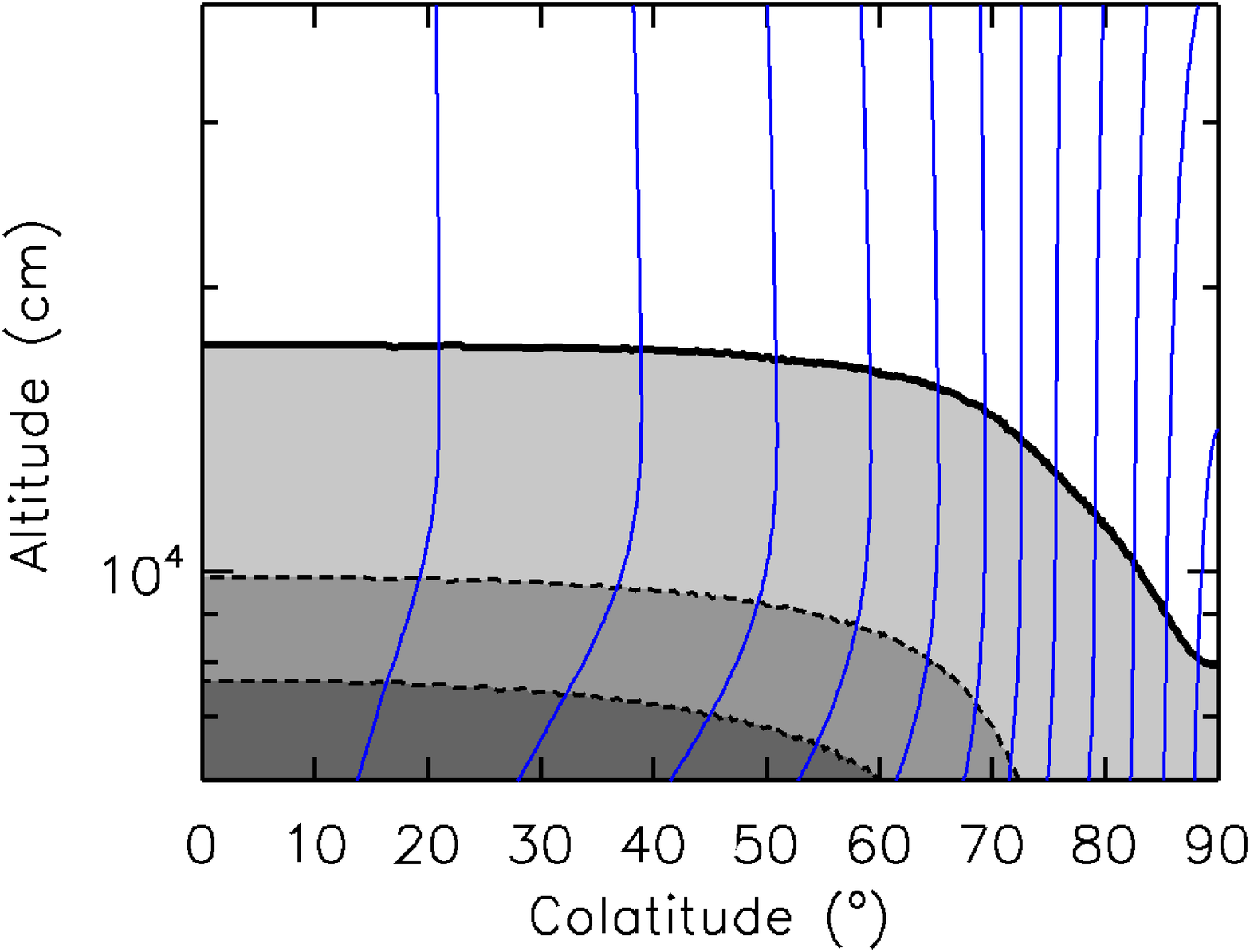}
}
\subfigure
{
	\includegraphics[width=84mm, height=64mm]{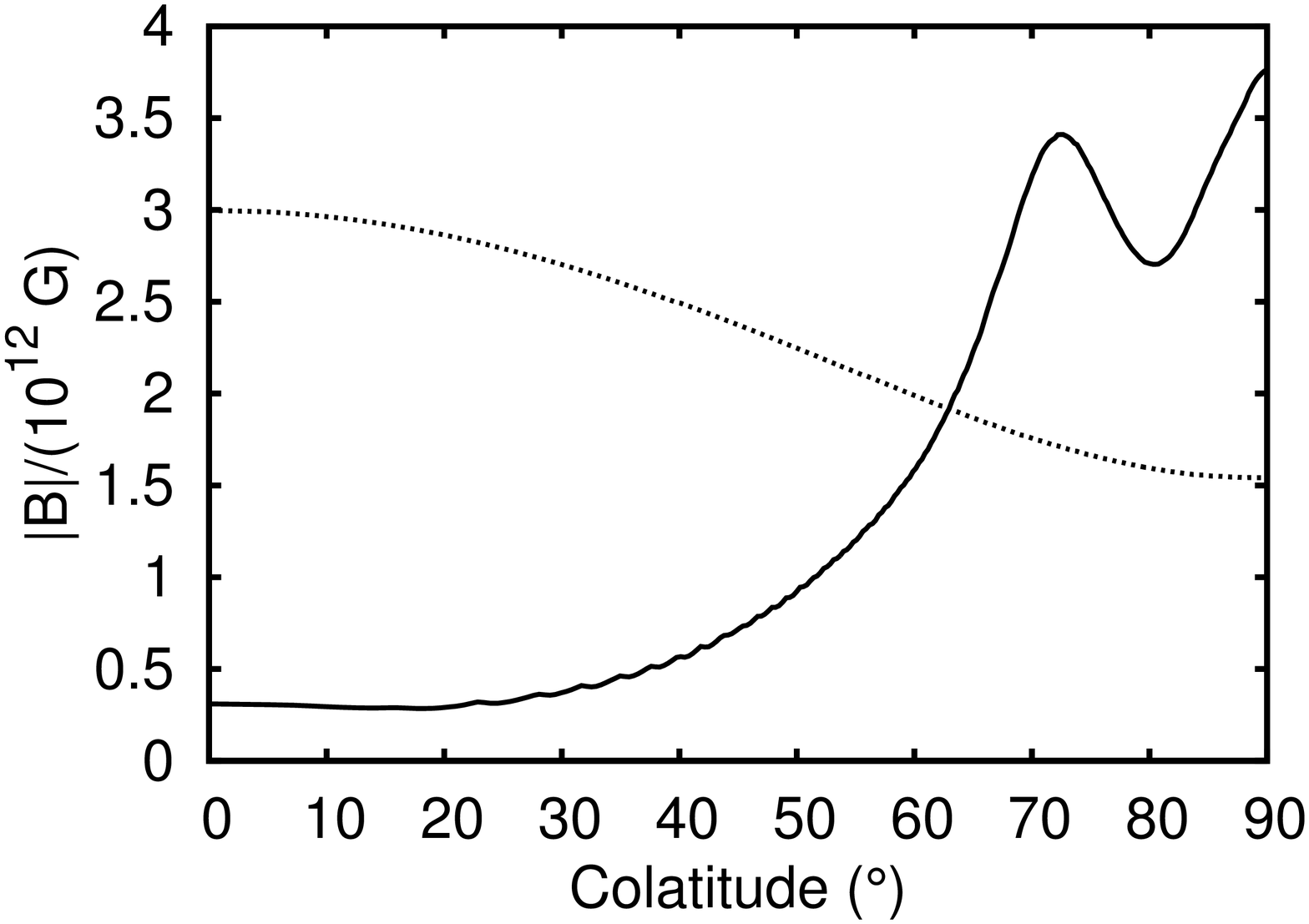}
}

\caption{Equilibrium hydromagnetic structure of an accretion mound with $M_{a} = M_{c}$, $\psi_{\ast} = 1.6 \times 10^{24} \ \mathrm{G} \ \mathrm{cm}^{2}$ 
and EOS model E \citep{priymak2011}. 
(\textit{Left panel}) Accretion mound surface $r=r_{m}(\theta)$ (thick solid curve) 
and magnetic field lines ($\psi$ contours; thin blue solid curves) in a meridional plane of constant $\phi$, with altitude $r-R_{\ast}$ and colatitude $\theta$ plotted 
on the vertical and horizontal axes respectively. Density contours are also drawn for $\rho=\eta \rho_{\mathrm{max}}$ with $\eta=0.2, 0.1$ (dashed curves) and 
shaded accordingly.
(\textit{Right panel}) Magnetic field strength $|\mathbf{B}|$, in units of $10^{12} \ \mathrm{G}$, on the mound surface (solid curve) as a function of colatitude, 
compared to an undistorted magnetic dipole with $\psi_{\ast} = 1.6 \times 10^{24} \ \mathrm{G} \ \mathrm{cm}^{2}$ (dotted curve).}
\label{fig:surface}
\end{minipage}
\end{figure*}

Fig. \ref{fig:surface} shows the output of the numerical GS solver for model E in \citet{priymak2011} and $\psi_{\ast} = 1.6 \times 10^{24} \ \mathrm{G} \ \mathrm{cm}^{2}$.
The pre-accretion magnetic field is distorted significantly when the accreted mass exceeds a characteristic critical value \citep{priymak2011}
\begin{equation}
\label{Mc}
M_{c} = 8.1 \times 10^{-8} (\psi_{\ast}/10^{24} \ \mathrm{G} \ \mathrm{cm}^{2})^{4/3} \ M_{\odot}.  
\end{equation}
Throughout the paper, we set $\psi_{\ast} = 1.6 \times 10^{24} \ \mathrm{G} \ \mathrm{cm}^{2}$, corresponding to a polar magnetic field 
strength of $10^{12.5} \ \mathrm{G}$, and take the neutron star mass to be $M_{\ast} = 1.4 M_{\odot}$.
The mountain in Fig. \ref{fig:surface} is constructed for $M_{a} = M_{c}$.
The thick black curve in the left panel denotes the outer surface of the mound $r=r_{m}(\theta)$
as a function of colatitude $\theta$, i.e. the contour $\rho[r_{m}(\theta), \theta]=0$. Dashed curves represent the density countours $\rho/\rho_{\mathrm{max}}=0.1, 0.2$, 
normalized by the maximum density $\rho_{\mathrm{max}}$ at $(r, \theta) = (R_{\ast}, 0)$. The thin blue curves describe contours of magnetic flux $\psi$, i.e. magnetic field lines.
The right panel of Fig. \ref{fig:surface} compares $|\mathbf{B}|$ at the mound surface as a function of $\theta$ (solid curve) with an undistorted dipole 
with the same hemispheric flux $\psi_{\ast}$ (dotted curve).
Magnetic burial reduces $|\mathbf{B}|$ by a factor of $\approx 6$ at the pole and increases $|\mathbf{B}|$ by a factor of $\approx 2$ at the equator 
relative to the dipole. In real systems, with $M_{a} \gg M_{c}$ typically, these factors are much higher.
The reader is reminded that the asphericity ostensibly apparent in Fig. \ref{fig:surface} is exaggerated by the choice of scale.
The height of the mound changes by $\approx 10^{4} \; \mathrm{cm}$ from $\theta \approx 60^{\circ}$ to $90^{\circ}$, over a lateral
distance of $\approx 5\times 10^{5} \; \mathrm{cm}$. Hence, to a good approximation, the surface is spherical, a property that we use
to simplify the cyclotron line calculations in Section \ref{sec:CRSF_model}.

For a mound like the one in Fig. \ref{fig:surface}, one has $\mu \approx 0.5 \mu_{i}$, much greater than observed in an LMXB.
For realistic $M_{a} \sim 0.1 M_{\odot}$ and $M_{a}/M_{c} \sim 10^{6}$, equation (\ref{Mc}) does predict low enough $\mu$,
but then the structure in Fig. \ref{fig:surface} needs to be modified: the equatorial belt is compressed into a narrower belt 
$80^{\circ} \la \theta \la 90^{\circ}$, with $B \gg 10^{13} \ \mathrm{G}$, and the field is buried at the pole, with $B \ll 10^{12} \ \mathrm{G}$
and $\mathbf{B}$ almost latitudinal. Unfortunately, numerical limitations (including artificial bubble formation; see Section \ref{sec:discussion})
of the GS method mean we cannot compute equilibria accurately for realistic $M_{a} \sim 0.1 M_{\odot}$; this challenging issue
besetting all the modeling to date \citep{brown1998,payne2004,mukherjee2012}.
Hence in this paper we do not claim that the predicted CRSF spectra are accurate but rather that they are indicative.
In reality, the CRSF is dominated by $\mathbf{B}$ at a narrow range of latitudes where $10^{12} \ \mathrm{G} \la B \la 10^{13} \mathrm{G}$,
halfway between as $B$ changes from $B \ll 10^{12} \ \mathrm{G}$ at pole to $B \gg 10^{12} \ \mathrm{G}$ at equator and this 
latitude range is narrower then the one that contributes to the CRSF spectra computed in this paper.
We reiterate this important caveat (and return to the issue of magnetic bubbles) in Sec \ref{sec:discussion}.

\subsection{Comparison with previous work}
\label{sec:comparison}
The mound calculations in this paper deliberately parallel those in \citet{mukherjee2012} in many ways, although our specific goal --- descriminating between
thermal and magnetic mountains --- differs from previous work. Both calculations assume magnetic line-tying at 
a hard surface at $r=R_{\ast}$ [i.e. both papers neglect sinking; cf. \citet{wette2010}], adopt a polytropic EOS, and assume that the mound is 
stable to ideal and resistive
hydromagnetic modes, in line with the findings of time-dependent MHD simulations \citep{payne2007, vigelius2008a, mukherjee2013a, mukherjee2013d}.
The main differences arise in the geometry, the boundary condition at the equator, and the way flux freezing is treated.
\citet{mukherjee2012} modeled the accretion mound within a polar cap of radius $1 \; \mathrm{km}$,
and assumed cylindrical symmetry. In comparison, we model the mound over an entire hemisphere. We therefore capture fully the magnetic tension
in the equatorially compressed magnetic field \citep{payne2004}, which increases $M_{c}$ by several orders of magnitude over previous estimates \citep{brown1998, litwin2001}.
\citet{mukherjee2012} adopted a strictly radial and uniform magnetic field at the surface, satisfying Dirichlet boundary 
conditions at the outer edge of the polar cap. 
We assume a dipolar magnetic field at the surface, 
and reflecting boundary conditions at the magnetic equator. 
Equatorial magnetic stresses 
stabilize the latter configuration against the growth of ideal-MHD ballooning modes for $M_{a} \la M_{c}$ \citep{vigelius2008a, mukherjee2012}; cf. \citet{litwin2001}. 
Finally, we solve self-consistently for $F(\psi)$ through equation (\ref{F_adiabatic}), whereas
\citet{mukherjee2012} prescribed the mound height profile as a function of flux 
without connecting it to a plausible pre-accretion $M(\psi)$ via flux-freezing.

\section{CRSF model}
\label{sec:CRSF_model}
The cyclotron spectrum emitted by a mound can be calculated in a few straightforward steps, once the magnetic field structure,
temperature distribution, and surface contour of the mound, and the orientation of the star relative to the observer, are specified. 
We follow closely the method pioneered by \citet{mukherjee2012}, who showed that cyclotron line formation occurs in a surface layer
$\sim 10 \ \mu \textrm{m}$ thick. The key formulas underpinning each step of the calculation are summarized 
briefly in sections \ref{sec:geometry} (geometry), \ref{sec:ray_tracing} (general relativistic light bending), \ref{sec:line_shape} (line depth 
and width), and \ref{sec:surface_flux} (surface flux distribution), with an emphasis on the new features that are added.
An example of a cyclotron spectrum is computed in Section \ref{sec:worked_example} to illustrate the key features, e.g. phase evolution,
which are studied further in Section \ref{sec:orientation}. 

\subsection{Geometry}
\label{sec:geometry}

\begin{figure*}
\begin{minipage}{170mm} 
\subfigure
{
	\includegraphics[width=84mm]{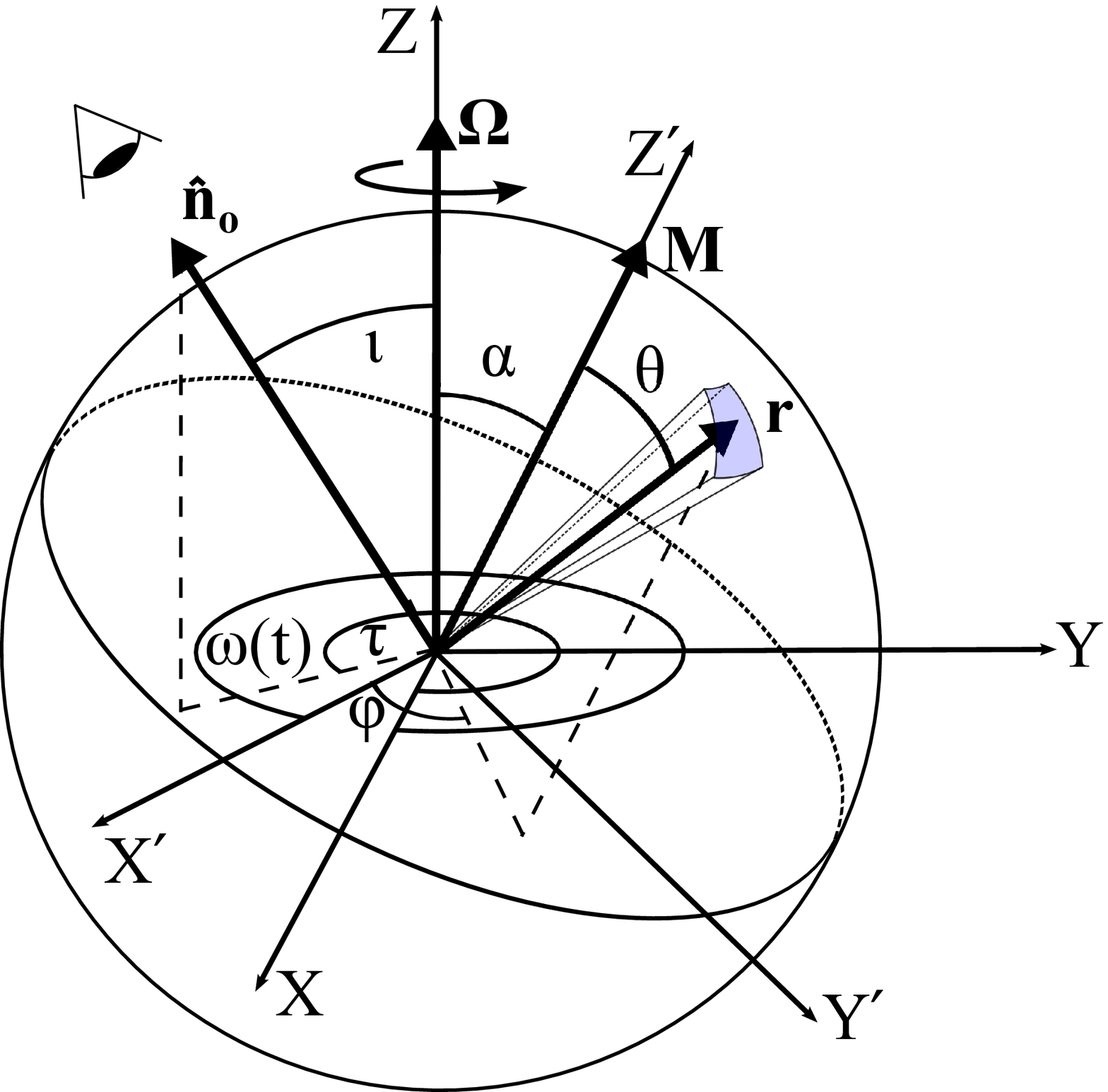}
}
\subfigure
{
	\includegraphics[width=84mm]{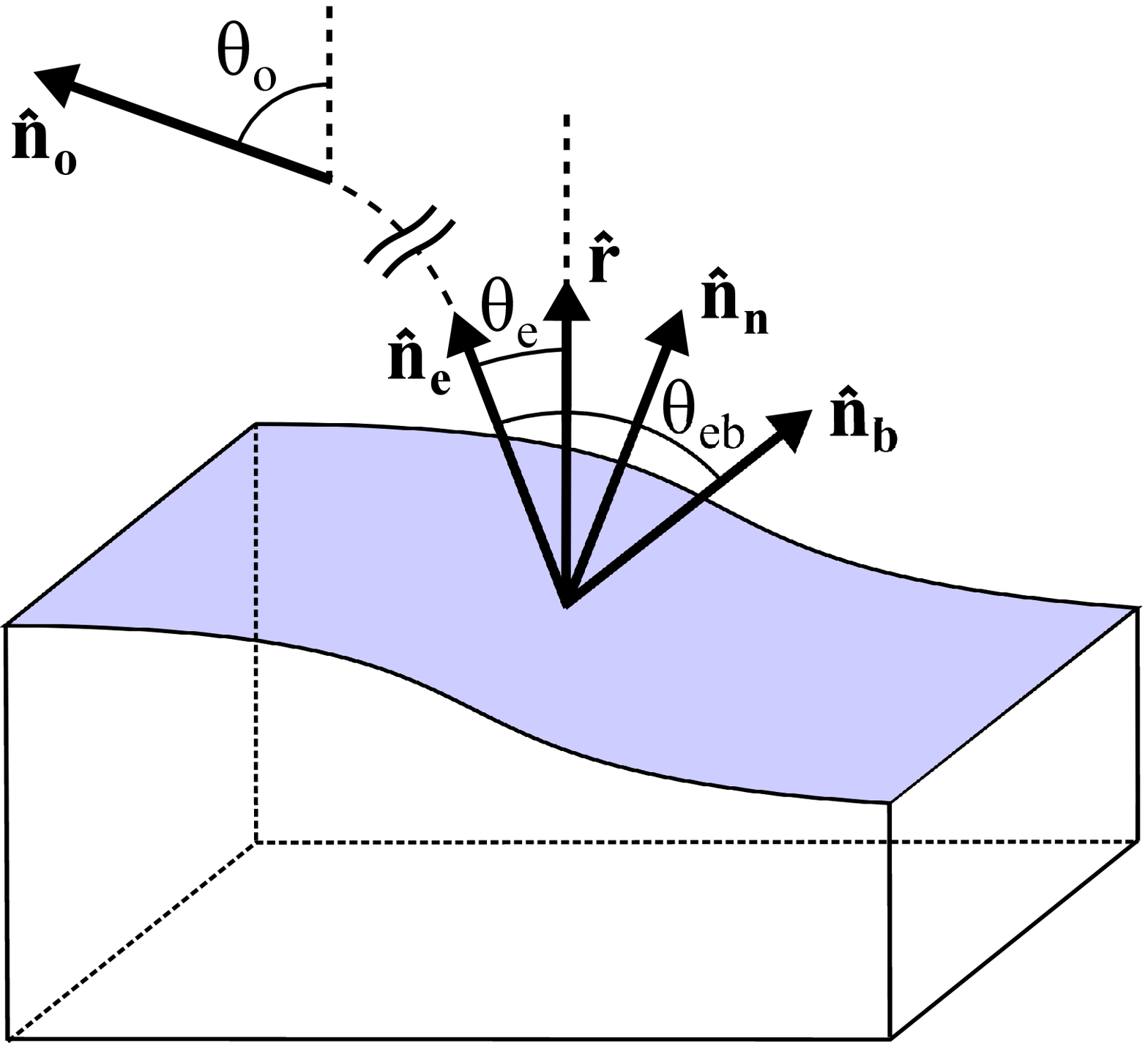}
}
\caption{(\textit{Left panel}) Orientation of the body frame $(X^{\prime}, Y^{\prime}, Z^{\prime})$ relative to the 
observer's frame $(X, Y, Z)$. The $Z^{\prime}$-axis is aligned with the magnetic symmetry axis $\mathbf{M}$ and inclined by an angle $\alpha$ 
with respect to $\mathbf{\Omega}$. The primed frame rotates about $\mathbf{\Omega}$. The  
position vector $\mathbf{r}$ to a surface element has colatitude $\theta$ and longitude $\phi$ in the primed frame. 
The line of sight $\hat{\mathbf{n}}_{o}$ makes angles $\iota$ and $\tau$ with the $Z$ axis and $X$ axis in the $X\text{--}Y$ plane.   
(\textit{Right panel}) Magnified view of the local surface of the accretion mound, showing the unit emission vector $\hat{\mathbf{n}}_{e}$, unit radial vector $\hat{\mathbf{r}}$, 
unit surface normal $\hat{\mathbf{n}}_{n}$, unit magnetic field vector $\hat{\mathbf{n}}_{b}$, and the angles between them. 
Due to gravitational light bending, photons emitted along $\hat{\mathbf{n}}_{e}$
follow a curved path (interrupted dashed curve) and are ultimately observed along $\hat{\mathbf{n}}_{o}$. In general, one has 
$\hat{\mathbf{n}}_{n} \cdot \hat{\mathbf{r}} \neq 1$ due to asphericity of the accretion mound, but $\hat{\mathbf{n}}_{n} \cdot \hat{\mathbf{r}} \approx 1$ 
holds to a good approximation. The vectors $\hat{\mathbf{r}}$, $\hat{\mathbf{n}}_{b}$, $\hat{\mathbf{n}}_{l}$, $\hat{\mathbf{n}}_{e}$ lie in the same plane
for axisymmetric GS equilibria (see Section \ref{sec:GS_equilibrium}).}
\label{fig:geometry_diagram}
\end{minipage}
\end{figure*}

In keeping with \citet{mukherjee2012}, we define a stationary (observer's) reference frame $(X, Y, Z)$ as drawn in the left panel of 
Fig. \ref{fig:geometry_diagram}, with origin $O$ at the centre of the 
neutron star and $Z$-axis coincident with the rotation axis $\mathbf{\Omega}$, which does not precess by assumption [cf. \citet{chung2008}]. 
In this frame, the colatitude and longitude of the line of sight (LOS) are $\iota$ and $\tau$ respectively, and the unit vector 
along the LOS is $\hat{\mathbf{n}}_{o} = (\sin \iota \sin \tau, \sin \iota \cos \tau, \cos \tau)$.

The accretion mound stands fixed on the surface of the star in a rotating reference frame $(X^{\prime}, Y^{\prime}, Z^{\prime})$, 
where $Z^{\prime}$ is always aligned with the magnetic symmetry axis $\mathbf{M}$, and $X$ coincides with $X^{\prime}$ at $t=0$. 
The GS calculation (neglecting Coriolis and centrifugal forces)\footnote{The ratio $\Lambda$ of the Coriolis force to the Lorentz force satisfies
$\Lambda \la 10^{-8}$ for circulatory settling motions with speed $v \ll 1 \ \mathrm{cm} \ \mathrm{s}^{-1}$ \citep{choudhuri2002} and $\Lambda \la 10^{-2}$
for a hypothetical zonal flow driven by accretion-induced temperature gradients [$v \la 10^{6} \ \mathrm{cm} \ \mathrm{s}^{-1}$ even during a thermonuclear
burst \citep{spitkovsky2002}], except perhaps near the pole, where the magnetic gradient is lower and $\Lambda \sim 1$ is possible. The centrifugal force is smaller than the Coriolis force by a factor $v/(R_{\ast}\Omega) \ll 1$.} is done in the primed frame
$(X^{\prime}, Y^{\prime}, Z^{\prime})$ in a spherical coordinate system $(r, \theta, \phi)$, 
where $\theta$ and $\phi$ are the colatitude and longitude with respect to $\mathbf{M}$, as in the left panel of Fig. \ref{fig:geometry_diagram}. 
The axis $\mathbf{M}$ makes an angle $\alpha$ with respect to $\mathbf{\Omega}$. 
The angle $\omega(t)$ between $X$ and $X^{\prime}$ in the $X-Y$ plane increases monotonically with $t$, as the star spins.
 
\subsection{Ray tracing}
\label{sec:ray_tracing}
Due to gravitational light bending, a ray intercepted by an observer along the LOS $\hat{\mathbf{n}}_{o}$ is 
emitted from a surface element on the mound at position $\mathbf{r}$ (shaded patch in the left panel of Fig. \ref{fig:geometry_diagram}) 
along a unit vector $\hat{\mathbf{n}}_{e}$,
which in general does not equal $\hat{\mathbf{n}}_{o}$. The right panel of Fig. \ref{fig:geometry_diagram} displays a magnified view of the emitting
surface element. 
The angles between $\hat{\mathbf{n}}_{o}$ and $\hat{\mathbf{r}}$, and 
$\hat{\mathbf{n}}_{e}$ and $\hat{\mathbf{r}}$, are $\theta_{o}$ and $\theta_{e}$ respectively. They can be related via the approximate 
formula \citep{beloborodov2002}
\begin{equation}
\label{gr_bending}
\cos \theta_{e} \approx u + (1 - u) \cos \theta_{o}, 
\end{equation}
where $u=R_{s}/r$, and $R_{s}$ is the Schwarzschild radius 
($R_{s} \approx 4.1 \ \mathrm{km}$ for a neutron star with a fiducial mass of $1.4 \ M_{\odot}$). 
This expression incorporates the effects of gravitational light bending
to an accuracy of $\sim 1 \%$ for typical neutron star parameters ($u \approx 1/3$).
We set $u=0.4$ throughout the paper.
As the unit vectors $\hat{\mathbf{n}}_{o}$, $\hat{\mathbf{n}}_{e}$, and $\hat{\mathbf{r}}$ lie in the same plane, we can write
\begin{equation}
\label{geometry}
\hat{\mathbf{n}}_{e} = \frac{\sin(\theta_{o} - \theta_{e})}{\sin \theta_{o}} \hat{\mathbf{r}} + \frac{\sin \theta_{e} }{\sin \theta_{o} } \hat{\mathbf{n}}_{o}.
\end{equation}

To calculate the contribution to the observed cyclotron line profile from the surface element at $\mathbf{r}$, we need to know the angle between
$\hat{\mathbf{n}}_{e}$ and the unit vector parallel to the local magnetic field, $\hat{\mathbf{n}}_{b} = \mathbf{B}/|\mathbf{B}|$.
The GS calculation in Section \ref{sec:accretion_mound} yields $\hat{\mathbf{n}}_{b}$ as well as the mound surface contour $r=r_{m}(\theta)$ in the 
frame $(X^{\prime}, Y^{\prime}, Z^{\prime})$. We therefore  
apply Euler rotations to $\hat{\mathbf{n}}_{o}$ about the $Z$ and $X$ axes by angles $-\omega(t)$ and $-\alpha$ respectively (see left panel of 
Fig. \ref{fig:geometry_diagram}) 
to bring it into the body frame $(X^{\prime}, Y^{\prime}, Z^{\prime})$ (i.e. $\hat{\mathbf{n}}_{o}$ depends on $t$ in the body frame). We then
use equation (\ref{geometry}) to calculate $\hat{\mathbf{n}}_{e}(r, \theta, \phi, t)$ at $\mathbf{r}$ in the body frame. 

\subsection{Line shape}
\label{sec:line_shape}
The angle $\theta_{eb}$ between $\hat{\mathbf{n}}_{e}$ and $\hat{\mathbf{n}}_{b}$ determines the line energy, width, and depth
of a CRSF feature. The central energy of a cyclotron line of order $n$ is given by
\begin{align}
\label{cyclotron_full}
E_{\mathrm{cyc,n}} = & m_{e} c^{2} (1 - u)^{1/2} \mathrm{cosec}^{2} \theta_{eb} \nonumber \\
 & \times \{ [1 + 2 n (B/B_{\mathrm{crit}}) \sin^{2} \theta_{eb}]^{1/2} - 1 \}, 
\end{align}
where $m_{e}$ is the electron rest mass, $c$ is the speed of light, 
$B_{\mathrm{crit}} = m^{2} c^{3}/(e \hbar) \approx 4.4\times10^{13} \: \mathrm{G}$ is the critical field strength, and 
the factor $(1 - u)^{1/2}$ arises from the gravitational redshift.
We only study the fundamental ($n=1$) cyclotron line in this paper. Since we have $B/B_{\mathrm{crit}} \la 10^{-1}$ at the mound surface, 
where most of the flux is emitted (see Fig. \ref{fig:surface}), we expand equation (\ref{cyclotron_full}) to first order in 
$B/B_{\mathrm{crit}}$ to give:
\begin{equation}
\label{cyclotron_approximation}
E_{\mathrm{cyc}}(\theta_{eb}, B) = E_{0} (1 - u)^{1/2} \left[ 1 - \Big( \frac{E_{0}}{1.02 \, \mathrm{MeV}} \Big) \sin^{2} \theta_{eb} \right], 
\end{equation}
with $E_{0} = 11.6 (B/10^{12} \ \mathrm{G})$ in keV. \citet{priymak2011} found $B/B_{\mathrm{crit}}$ as 
high as $\sim 10^{2}$ in parts of the accretion mound, but these regions lie well beneath the thin emitting skin at the surface
for all relevant EOS models.

We assume isotropic emission at each point on the surface [i.e. $F(\theta_{e}, E, \theta_{eb}) = F(E, \theta_{eb})$], 
and simplify the emitted spectrum in the vicinity of $E_{\mathrm{cyc}}$ as a flat continuum (of unit flux)
modulated by the CRSF [cf. \citet{coburn2002}], which is assumed to be a Gaussian in absorption:
\begin{equation}
\label{cyclotron_line}
F(E, \theta_{eb}) = 1 - [1 - D(\theta_{eb})] \exp \Big[ - \frac{2.8 (E - E_{\mathrm{cyc}})^{2}}{F(\theta_{eb})^{2}} \Big]. 
\end{equation}
Here, $D(\theta_{eb})$ is the line depth, and $F(\theta_{eb})$ is the full width half maximum.
Equation (\ref{cyclotron_line}) is not strictly correct, because a typical CRSF profile is more complicated than a 
Gaussian \citep{kreykenbohm2005, schonherr2007}, but we follow \citet{mukherjee2012} in adopting the Gaussian approximation.

The line width $F(\theta_{eb})$ and depth $D(\theta_{eb})$ are obtained directly from Fig. 8 of \citet{schonherr2007}, for a slab 1-0 geometry 
(plane parallel homogeneous layer illuminated from below). A slab is a fair approximation
to a small element on the mound surface, whereas a cylindrical geometry is more appropriate for the accretion column as a whole.
The spectral contribution arising from fan-beam emission from a radiation-dominated accretion column should not be ignored for supercritical
accretors with accretion luminosities comparable to the Eddington luminosity \citep{basko1976, nelson1993, becker2012}.
In this paper, for simplicity, we model the cyclotron spectra exclusively from subcritical accretors, 
which are often the subject of gravitational-wave studies, and do not consider cylindrical geometry any further.
The slab 1-1 geometry (plane parallel homogeneous layer illuminated at the mid plane) yields qualitatively similar results to 1-0, because $F(\theta_{eb})$ and $D(\theta_{eb})$ have similar functional forms
in the two cases [see Fig. 8 of \citet{schonherr2007}]. Given the inability of current observations to discriminate between 1-0 and 1-1, 
we do not consider the 1-1 geometry further.

\subsection{Surface flux}
\label{sec:surface_flux}
We divide the neutron star surface into equal-area elements $\Delta A_{\theta_{i}, \phi_{j}}$ at latitude $\theta_{i}$ and longitude $\phi_{j}$.
The observed spectrum is constructed by numerically integrating the emitted 
spectrum $F(E, \theta_{eb})$ from equation (\ref{cyclotron_line}) over the observable 
($\hat{\mathbf{n}}_{e} \cdot \hat{\mathbf{n}}_{n} > 0$) 
grid elements and multiplying by the projected area perpendicular to the LOS:
\begin{equation}
F_{tot}(E) = \sum\limits_{\theta_{i}, \phi_{j}} F[E, \theta_{eb}(\theta_{i}, \phi_{j})] \Delta A(\theta_{i}, \phi_{j}) \cos \theta_{e}.
\label{numerical_integration}
\end{equation}
For plotting purposes, we normalize the integrated spectrum from 
equation (\ref{numerical_integration}) to give unit peak flux.
Strictly speaking, the surface of the accretion mound is not spherical. Hence the unit normal $\hat{\mathbf{n}}_{n}$ of any surface element
does not equal the unit position vector $\hat{\mathbf{r}}$ of that element (see right panel of Fig. \ref{fig:geometry_diagram}). In this paper, 
however, we make the approximation 
$\hat{\mathbf{n}}_{n} \approx \hat{\mathbf{r}}$, which is accurate everywhere [maximum $\sim 1 \%$ error at $\theta \approx \pi/2$; see Fig. \ref{fig:surface}],
as $\hat{\mathbf{n}}_{n}$ is hard to calculate accurately from the numerical GS output.

\subsection{Worked example}
\label{sec:worked_example}

\begin{figure*}
\begin{minipage}{170mm} 
\subfigure
{
	\includegraphics[width=84mm]{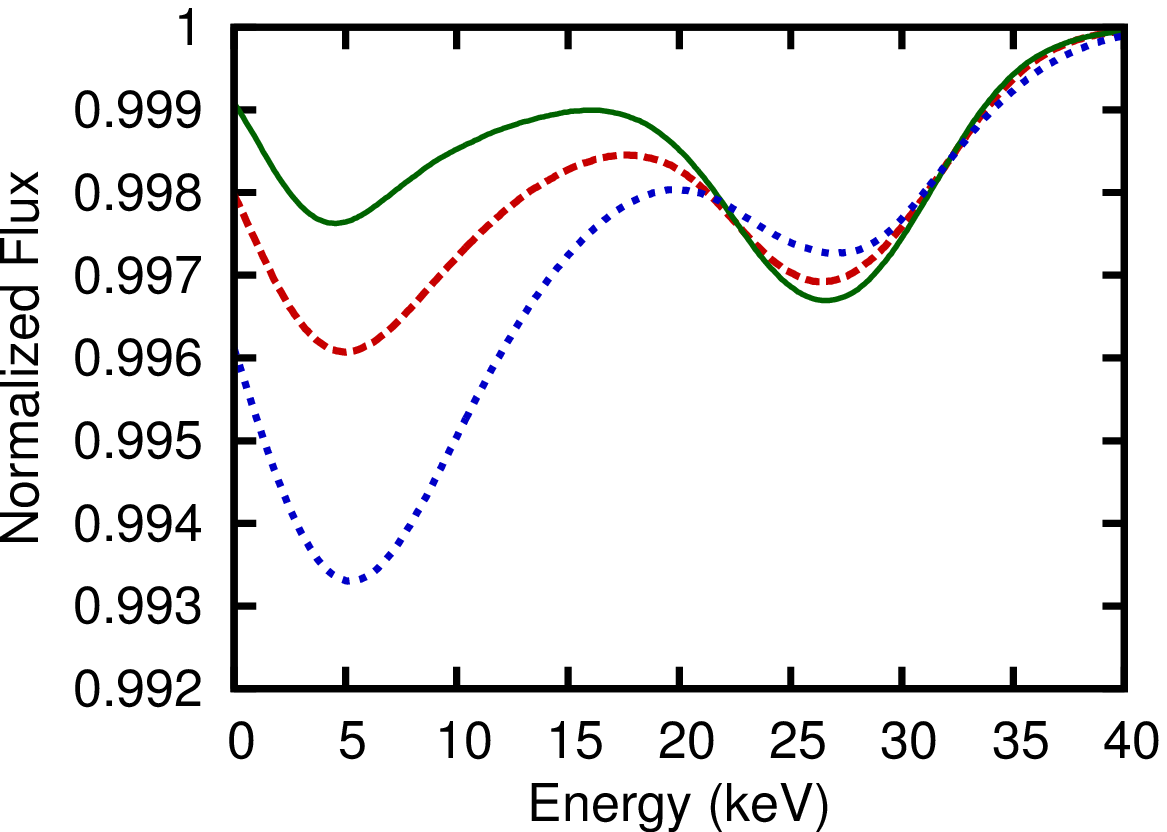}
}
\subfigure
{
	\includegraphics[width=84mm]{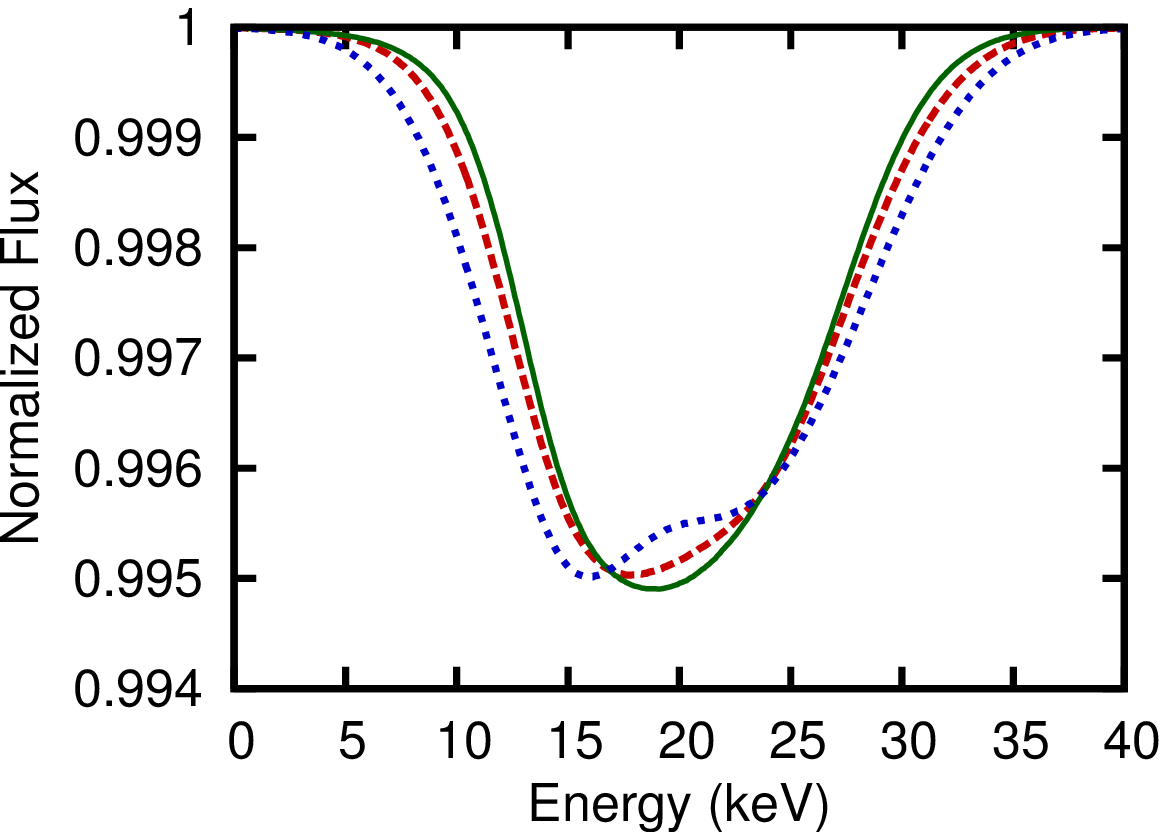}
}
\caption{Phase-dependent normalized CRSF spectrum with $(\iota, \alpha) = (\pi/4, \pi/6)$, $\psi_{\ast} = 1.6 \times 10^{24} \ \mathrm{G} \ \mathrm{cm}^{2}$, 
and $u=0.4$. (\textit{Left panel}) Magnetic mountain with $M_{a} = M_{c}$ and EOS model E \citep{priymak2011}. 
(\textit{Right panel}) Undisturbed magnetic dipole with $M_{a} = 0$.
The three curves in each panel correspond to three rotational phases: $\omega=0$ (dashed red), $2\pi/3$ (solid green), and $4\pi/3$ (dotted blue).}
\label{fig:mound_dipole_comparison}
\end{minipage}
\end{figure*}

\begin{figure*}
\begin{minipage}{170mm} 
\subfigure
{
	\includegraphics[width=170mm]{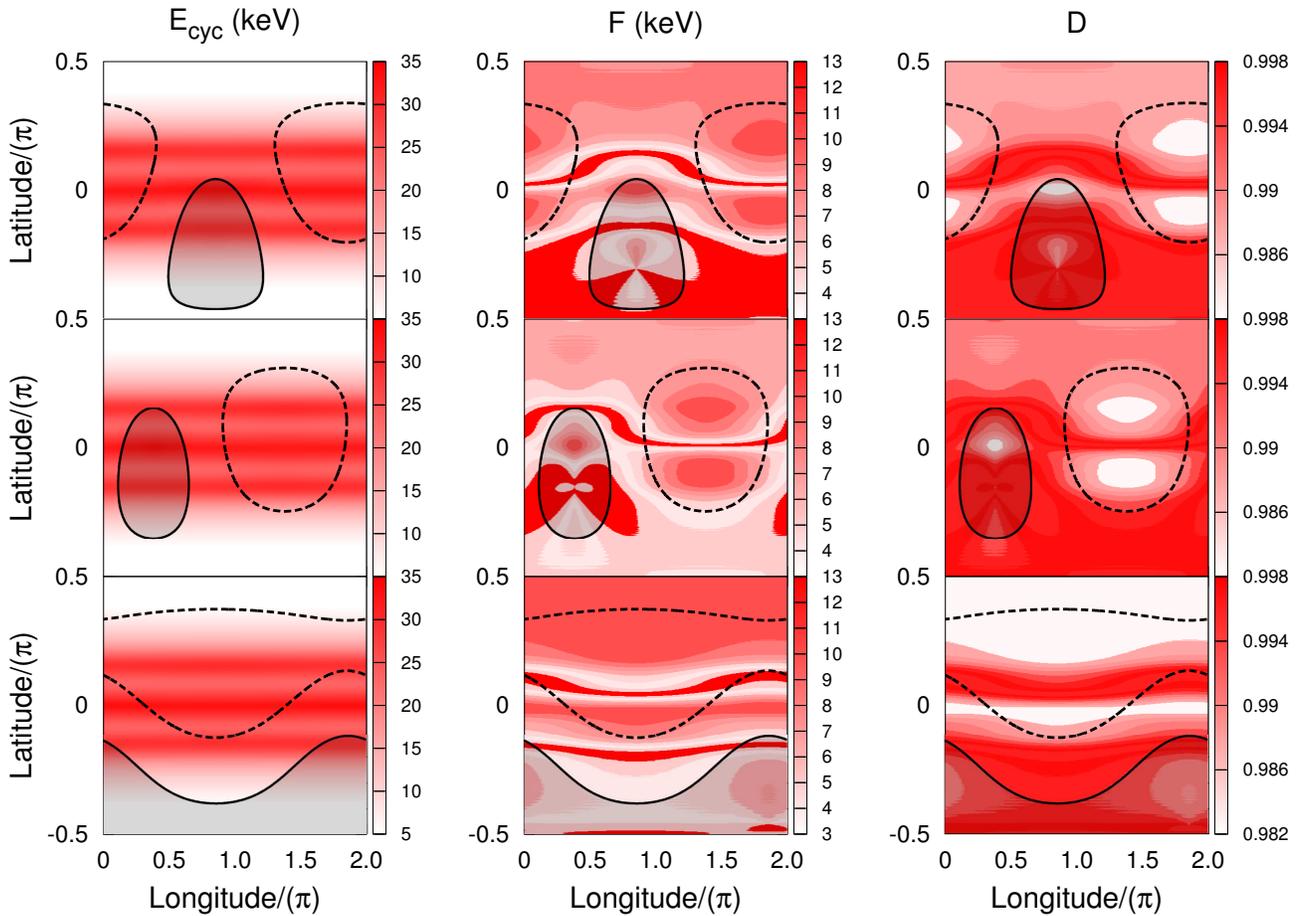}
}
\caption{Surface variation of CRSF parameters $E_{\mathrm{cyc}}$ (line energy in keV), $F$ (full width half maximum in keV), and $D$ (depth in normalized flux units) 
(left to right columns) as functions of rotational phase $\omega=0, \ 2\pi/3, \ 4\pi/3$ (top to bottom rows) 
with $(\iota, \alpha) = (\pi/4, \pi/6)$, $\psi_{\ast} = 1.6 \times 10^{24} \ \mathrm{G} \ \mathrm{cm}^{2}$, $u=0.4$, $M_{a} = M_{c}$, 
and EOS model E \citep{priymak2011}.
The vertical and horizontal axes represent latitude $\pi/2 - \theta$ and 
longitude $\phi$ respectively, in units of $\pi$ radians. The magnitude of the plotted parameters is described by the red shading, with minimum/maximum values
plotted as light/dark respectively. Regions on the surface which do not contribute to the observed
spectrum ($\mathbf{\hat{n}}_{e} \cdot \mathbf{\hat{r}} \leq 0$) and which dominate the surface-integrated cyclotron spectrum, 
i.e. $(\mathbf{\hat{n}}_{e} \cdot \mathbf{\hat{r}}) \mathrm{d}A \geq \mathrm{max}[(\mathbf{\hat{n}}_{e} \cdot \mathbf{\hat{r}}) \mathrm{d}A]/2$,
are enclosed by solid and dashed black curves respectively.}
\label{fig:surface_1}
\end{minipage}
\end{figure*}

\begin{figure*}
\begin{minipage}{170mm} 
\subfigure
{
	\includegraphics[width=170mm]{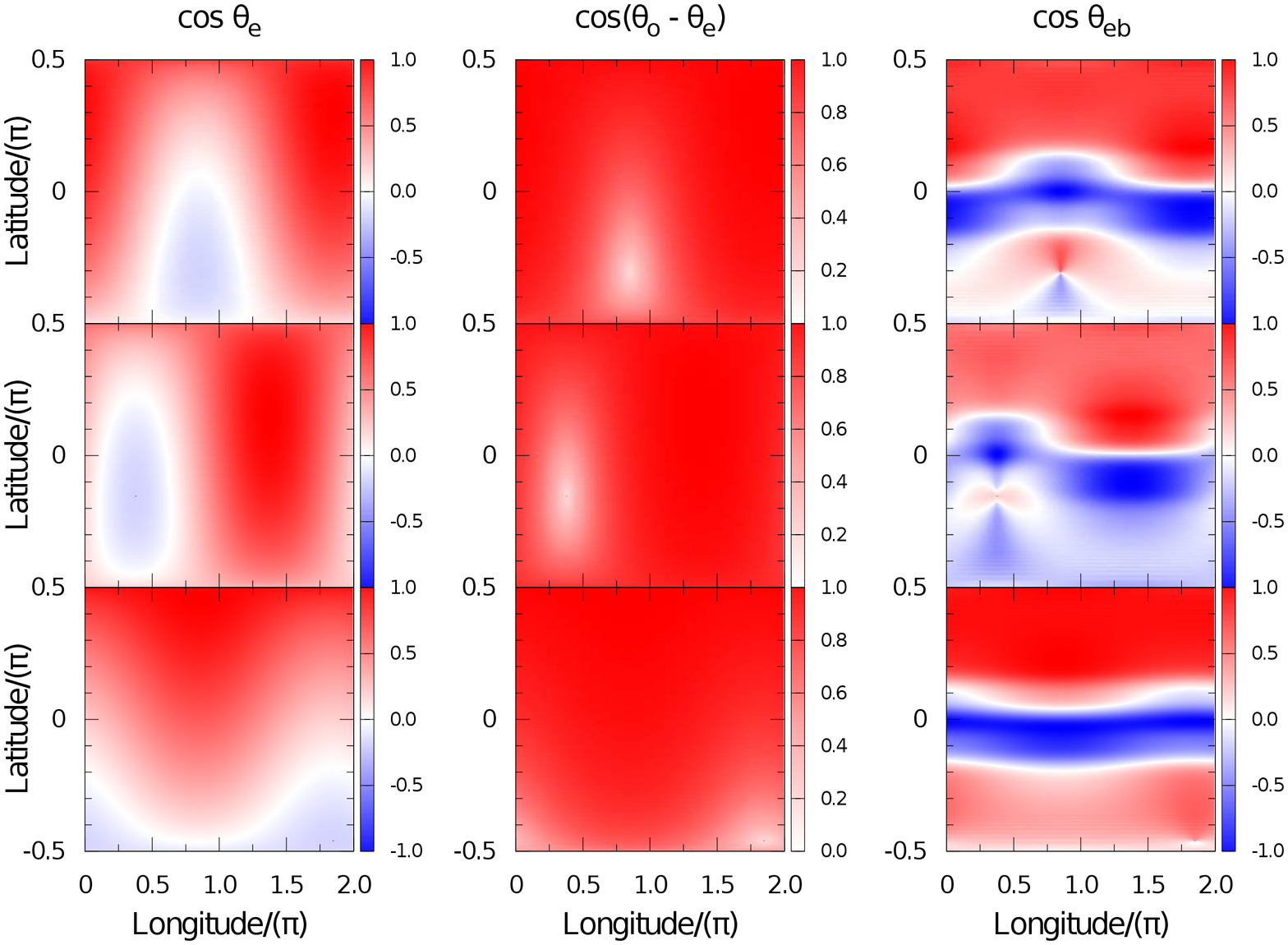}
}
\caption{Surface variation of direction cosines $\cos \theta_{e}$, $\cos (\theta_{o} - \theta_{e})$, and $\cos \theta_{eb}$ (left to right columns) 
as functions of rotational phase $\omega=0, \ 2\pi/3, \ 4\pi/3$ (top to bottom rows) with 
$(\iota, \alpha) = (\pi/4, \pi/6)$, $\psi_{\ast} = 1.6 \times 10^{24} \ \mathrm{G} \ \mathrm{cm}^{2}$, $u=0.4$, $M_{a} = M_{c}$, 
and EOS model E \citep{priymak2011}.
The vertical and horizontal axes represent latitude $\pi/2 - \theta$ and 
longitude $\phi$ respectively, in units of $\pi$ radians. The magnitude of the plotted direction cosines is described by the red shading, 
with minimum/maximum values plotted as light/dark respectively.}
\label{fig:surface_2}
\end{minipage}
\end{figure*}

Figure \ref{fig:mound_dipole_comparison} (left panel) shows the spectrum emitted by the accretion mound of Fig. \ref{fig:surface} when it is observed 
from the vantage point $\iota=\pi/4, \ \tau=0$ with magnetic inclination $\alpha=\pi/6$. 
The spectrum is plotted at three phases ($\omega=0, \ 2\pi/3, \ 4\pi/3$).
Also shown in the right panel is the spectrum at the same phases before accretion ($M_{a} = 0$), when the magnetic field is an undisturbed dipole. 
In the next two figures, to aid interpretation, we compute the line parameters $E_{\mathrm{cyc}}$, $F$, and $D$ (Figure \ref{fig:surface_1}) and direction 
cosines $\hat{\mathbf{n}}_{e} \cdot \hat{\mathbf{r}}$, $\hat{\mathbf{n}}_{e} \cdot \hat{\mathbf{n}}_{o}$, and $\hat{\mathbf{n}}_{e} \cdot \hat{\mathbf{n}}_{b}$ 
(Figure \ref{fig:surface_2}) as functions of colatitude $\theta$ and longitude $\phi$ in the body frame for the same three rotational phases 
$\omega=0, \ 2\pi/3, \ 4\pi/3$ (top to bottom rows).

The cyclotron spectrum from an undistorted magnetic dipole (right panel of Fig. \ref{fig:mound_dipole_comparison}) exhibits a single absorption trough at 
$E_{\mathrm{cyc}} \approx 20 \ \mathrm{keV}$. It is approximately phase invariant, the line properties change by $\Delta E_{\mathrm{cyc}} \approx 5 \ \mathrm{keV}$, 
$\Delta F \approx 3 \ \mathrm{keV}$, and $\Delta D \approx 10^{-4}$ over one rotation period.
In comparison, the spectrum generated from the surface of the accretion mound (left panel of Fig. \ref{fig:mound_dipole_comparison})
displays two distinct CRSFs at $E_{\mathrm{cyc},1} \approx 5 \ \mathrm{keV}$ and $E_{\mathrm{cyc},2} \approx 27 \ \mathrm{keV}$ with 
$\Delta E_{\mathrm{cyc},1} \approx 1 \ \mathrm{keV}$, $\Delta F_{1} \approx 5 \ \mathrm{keV}$, $\Delta D_{1} \approx 5 \times 10^{-3}$, 
and $\Delta E_{\mathrm{cyc},2} \approx 2 \ \mathrm{keV}$, $\Delta F_{2} \approx 5 \ \mathrm{keV}$, $\Delta D_{2} \approx 5 \times 10^{-4}$.
The solid black curves in Fig. \ref{fig:surface_1} and red shading in the left column of Fig. \ref{fig:surface_2} show that more than half 
the surface contributes to the observed spectrum ($\hat{\mathbf{n}}_{e} \cdot \hat{\mathbf{r}} > 0$) 
due to gravitational light bending.
The angle $\theta_{o} - \theta_{e}$ increases as $\hat{\mathbf{r}}$ tilts away from $\hat{\mathbf{n}}_{o}$
[see equations (\ref{gr_bending}) and (\ref{geometry})] 
and is rarely greater than $\approx 60^{\circ}$ (see middle column of Fig. \ref{fig:surface_2}). 
Although the unit magnetic field vectors $\hat{\mathbf{n}}_{b}$ are always axisymmetric (in this specific GS geometry), 
the emission unit vectors $\hat{\mathbf{n}}_{e}$ are not axisymmetric for non-trivial $\hat{\mathbf{n}}_{o}$ and $\mathbf{M}$ orientations, as $\theta_{eb}$ depends on $\phi$ and $t$ 
(rightmost column of Fig. \ref{fig:surface_2}). As a corollary, the region on the surface which dominates the flux,
i.e. $(\mathbf{\hat{n}}_{e} \cdot \mathbf{\hat{r}}) \mathrm{d}A \geq \mathrm{max}[(\mathbf{\hat{n}}_{e} \cdot \mathbf{\hat{r}}) \mathrm{d}A]/2$ 
(see dashed black curves in Fig. \ref{fig:surface_1}), is similarly non-axisymmetric
and phase-dependent. 

The appearance of two narrow absorption features from a mountain is explained by the contribution of both the compressed equatorial magnetic 
field (at $\approx 27 \ \mathrm{keV}$ from latitude $0 \la |\pi/2 - \theta| \la 0.3\pi$)
and the diluted polar magnetic field (at $\approx 5 \ \mathrm{keV}$ from latitudes $0.3\pi \la |\pi/2 - \theta| \la \pi$) (leftmost column of Fig. \ref{fig:surface_1}).
The line at $E_{\mathrm{cyc},1}$ is the dominant absorption feature for $2\pi/3 \leq \omega \leq 2\pi$ because a larger fraction of the polar region
contributes to the surface-integrated flux for $2\pi/3 \la \omega \la 4\pi/3$ than for $0 \la \omega \la 2\pi/3$. 
You can see this in Fig. \ref{fig:surface_2} by comparing the areas enclosed by the dashed black curves in the top, middle, and bottom panels.
Here the polar region is that where the distorted magnetic field is approximately parallel to the emission direction
($\hat{\mathbf{n}}_{e} \cdot \hat{\mathbf{n}}_{b} \approx 1$); see leftmost and rightmost columns of Fig. \ref{fig:surface_2}.

The line depth $D$ varies substantially over one full rotation period 
(rightmost column of Fig. \ref{fig:surface_1} and leftmost column of Fig. \ref{fig:surface_2}),
especially for the line at $E_{\mathrm{cyc},1}$ ($\Delta D_{1} \approx 5 \times 10^{-3}$). 
The line $E_{\mathrm{cyc},1}$ is generated primarily in the polar regions, and $\theta_{eb}$ in this
region varies with rotational phase (see rightmost column of Fig. \ref{fig:surface_2}).
Since $D$ and $F$ are directly related to $\cos \theta_{eb}$ at each emission point,
in such a way that there exists a monotonic correlation between them [see middle and rightmost columns of Fig. \ref{fig:surface_1} and \citet{schonherr2007}],
we expect to see some part of that correlation survive after integrating over the observable area ($\hat{\mathbf{n}}_{e} \cdot \hat{\mathbf{r}} > 0$),
so that the line is widest when it is deepest, and vice versa.

\section{Orientation}
\label{sec:orientation}

\begin{figure*}
\begin{minipage}{175mm}
\subfigure
{
	\includegraphics[width=175mm]{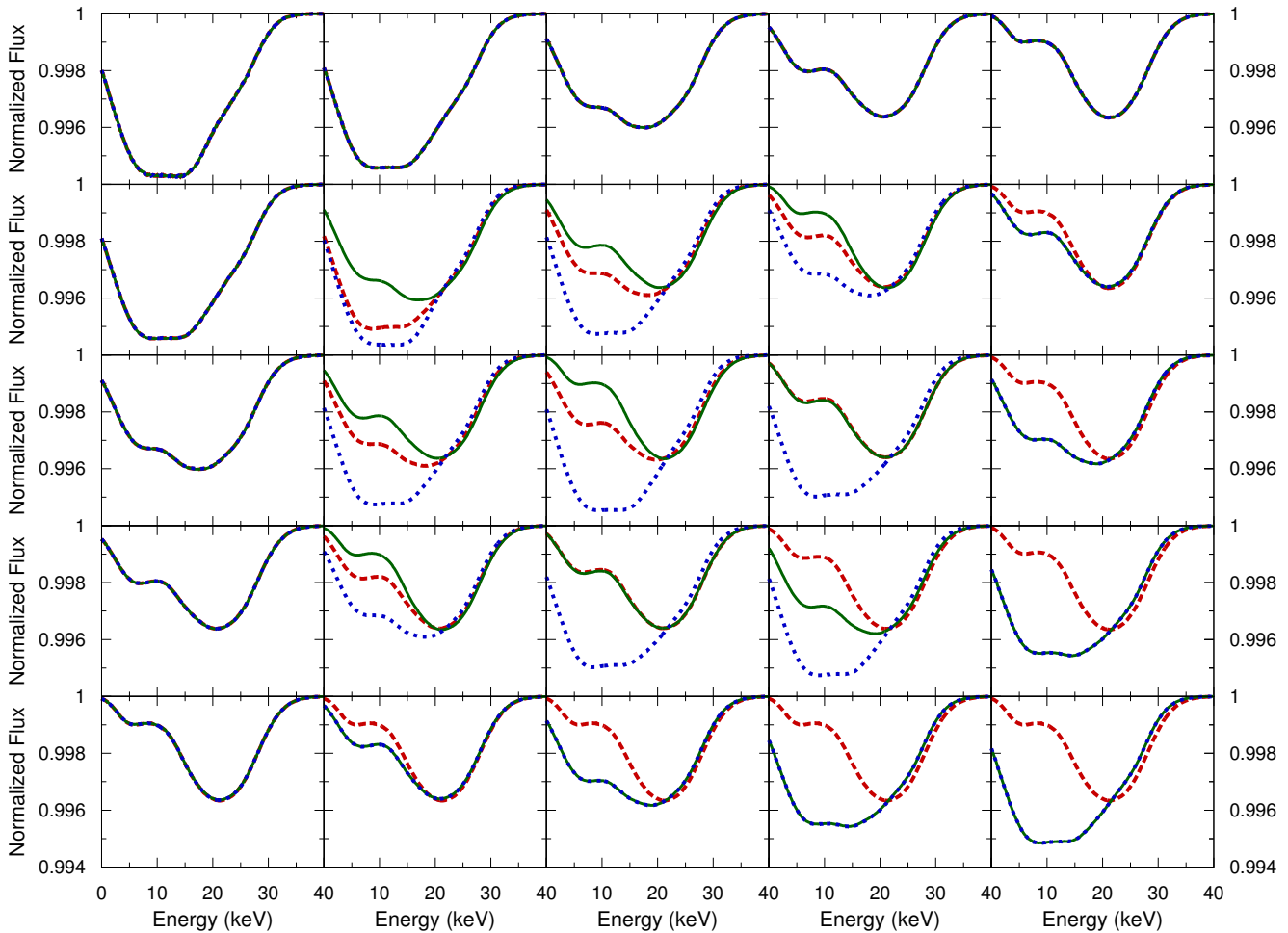}
}
\caption{Phase-resolved normalized CRSF spectra for a magnetic mountain with $M_{a} = 0.75 M_{c}$, 
$u=0.4$, $\psi_{\ast} = 1.6 \times 10^{24} \ \mathrm{G} \ \mathrm{cm}^{2}$, and EOS model E \citep{priymak2011}, covering a range of observer inclination angles 
$\iota = 0$, $\pi/8$, $\pi/4$, $3\pi/8$, $\pi/2$ (top to bottom rows)
and magnetic inclination angles $\alpha=0$, $\pi/8$, $\pi/4$, $3\pi/8$, $\pi/2$ (left to right column). 
In each panel, three phases ($\omega=0$, $2\pi/3$, $4\pi/3$) are plotted as long-dashed red, solid green, and dotted blue curves respectively.}
\label{fig:spectra(i,alpha)}
\end{minipage}
\end{figure*}

The detailed shape of a CRSF and its variation with rotational phase depend on the magnetic inclination $\alpha$ and the observer's inclination $\iota$.
In this section we show that phase-resolved CRSF observations can constrain $\alpha$ and $\iota$ (without fixing them uniquely; e.g.
the spectrum is unchanged when switching $\alpha$ and $\iota$). As well as assisting with studies of surface structure, this
capability may play a role in future in reducing the 
parameter space for computationally expensive gravitational wave searches. Constraining $\alpha$ and $\iota$ is the first step in discriminating
between thermal and magnetic mountains in various systems, as described in Section \ref{sec:discriminating_mountains}.

The spectrum emitted by a mound of mass $M_{a} = 0.75 M_{c}$ is displayed in Figure \ref{fig:spectra(i,alpha)}, for different observer 
orientations ($\iota=0,\pi/4,3\pi/8,\pi/2$, top to bottom rows),
inclination angles ($\alpha=0,\pi/4,3\pi/8,\pi/2$, top to bottom rows), and rotational phases $\omega = 0, 2\pi/3, 4\pi/3$ 
(long-dashed red, solid green, and dotted blue curves respectively). 
We set $\tau=0$ for convenience; $\tau$ simply introduces a phase offset.

For $\alpha = 0$ (left column) and $\iota = 0$ (top row), the spectrum is independent of phase.
Emission from the region $55^{\circ} \la \theta \la 65^{\circ}$ 
dominates the overall spectrum, weighted by the factor $\cos \theta_{e}(\theta) \, \mathrm{d}A(\theta)$.
The relatively wide and deep CRSF at $(\iota, \alpha) = (0,0)$ separates into two distinct absorption troughs 
as $\iota$ or $\alpha$ increase, and the troughs simultaneously become shallower. 
Emission from the polar cap dominates the spectrum at $(\iota, \alpha) = (0,0)$.
The equatorial belt $(70^{\circ} \la \theta \la 90^{\circ})$ contributes more, as either $\iota$ or $\alpha$ increase.
Since $|\mathbf{B}|$ is greater in the equatorial belt than at the pole, the deeper trough shifts to higher energies, while the shallower trough shifts
to lower energies, as accreted matter spreads equatorward and $|\mathbf{B}|$ at the pole weakens (see Fig. \ref{fig:surface}). 
The line depth falls with $\iota$ or $\alpha$ because $\theta_{eb}$ increases on average over the region 
$\hat{\mathbf{n}}_{e} \cdot \hat{\mathbf{r}} > 0$ [see Fig. \citet{schonherr2007}].

When neither $\iota$ nor $\alpha$ are near zero, it is harder to draw definite conclusions about the viewing geometry from observations.
Nevertheless, some constraints can be deduced. If either $\iota$ or $\alpha$ approaches $\pi/2$, the spectrum transitions between two distinct states within
one period, at twice the rotational frequency. In contrast, for $\pi/8 \la \iota, \alpha \la 3\pi/8$, the spectrum transitions between three states within
one period, if sampled at three equally spaced rotational phases. This information, albeit incomplete, is potentially useful for
surface magnetic field reconstruction and for optimizing gravitational-wave searches.

\section{Accreted Mass}
\label{sec:accreted_mass}

\begin{figure*}
\begin{minipage}{175mm}
\subfigure
{
	\includegraphics[width=175mm]{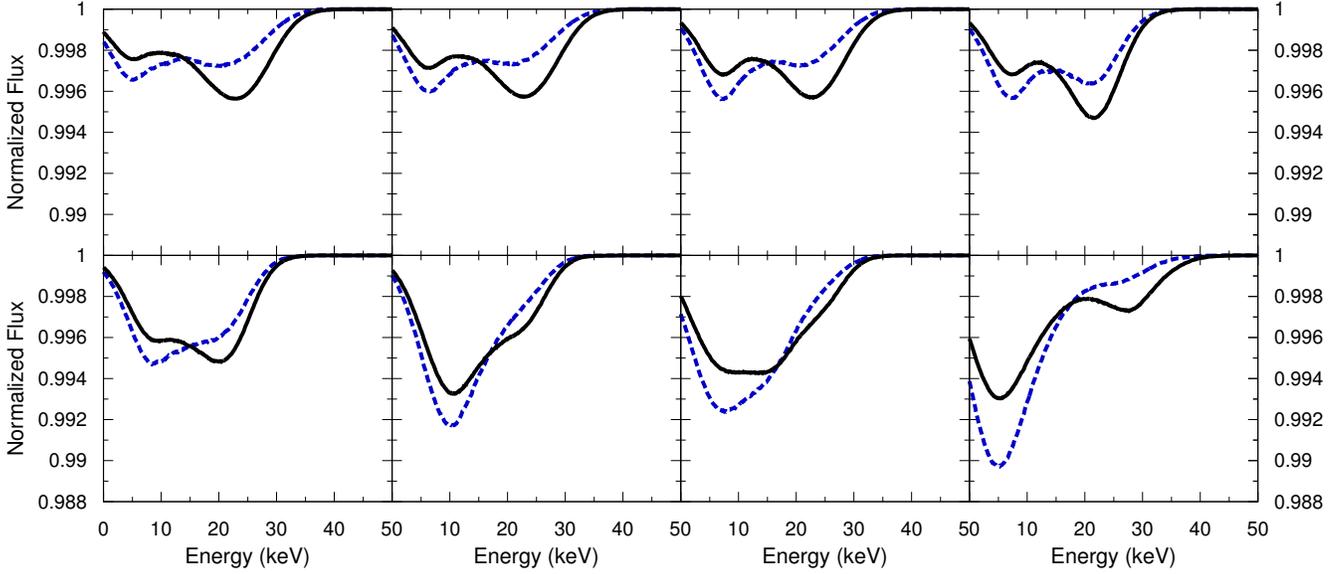}
}
\caption{CRSF spectra for a magnetic mountain in the aligned configuration $(\iota, \alpha) = (0,0)$ (i.e. no phase dependence), with $u=0.4$, 
$\psi_{\ast} = 1.6 \times 10^{24} \ \mathrm{G} \ \mathrm{cm}^{2}$, and EOS model E \citep{priymak2011}, as a function of accreted mass 
$M_{a}/M_{c} = 10^{-4}$, $10^{-3}$, $10^{-2}$, $10^{-1}$ (top row, left to right column), and 
$M_{a}/M_{c} = 0.25$, $0.5$, $0.75$, $1.0$ (bottom row, left to right column). 
In each panel, the solid and dashed curves correspond to uniform and nonuniform, magnetic-pressure modified surface temperature distributions 
respectively.}
\label{fig:accreted_mass}
\end{minipage}
\end{figure*}

The accreted mass $M_{a}$ in observed systems cannot be measured directly and is usually estimated via evolutionary studies \citep{podsiadlowski2002, pfahl2003}. 
$|\mathbf{B}|$ and the ellipticity $\epsilon$ of the star are partly determined by $M_{a}$ 
(with lesser contributions from the specific accretion geometry and the EOS); both are important properties of the accretion mound's structure.
In addition, $\epsilon$ determines the characteristic strain $h_{0}$ of the gravitational wave signal.
It is therefore useful to both magnetic field and gravitational wave studies to examine how $M_{a}$ affects the cyclotron spectrum.
A detection of CRSFs in an accreting neutron star can, in principle, constrain $M_{a}$ and hence help predict $\epsilon$ and $h_{0}$.

Fig \ref{fig:accreted_mass} presents the spectrum generated by EOS model E (solid black curve) for a mound with constant surface 
temperature $T = 3.5\times10^{7} \ \mathrm{K}$ \citep{schonherr2007}
for an aligned configuration with $(\iota, \alpha) = (0, 0)$ and hence no phase dependence, $u=0.4$, and 
$\psi_{\ast} = 1.6 \times 10^{24} \mathrm{G} \ \mathrm{cm}^{2}$. The spectrum is plotted in eight panels as a function of $M_{a}$, 
with $M_{a}/M_{c} = 10^{-4}$, $10^{-3}$, $10^{-2}$, $10^{-1}$ along the top row (left to right),
and $M_{a}/M_{c} = 0.25$, $0.5$, $0.75$, $1.0$ along the bottom row (left to right).

As accretion proceeds, the magnetic field at the surface of the mound is increasingly distorted.
For $M_{a} \la 10^{-1} \ M_{c}$, both troughs deepen as $M_{a}$ increases, 
because the angle $\theta_{eb}$ increases over the observed area, as $\mathbf{B}$ is compressed laterally.
For $M_{a} \ga 10^{-1} \ M_{c}$, the monotonic trend ceases, and the line at $E_{\mathrm{cyc},1}$ becomes deeper than the line at 
$E_{\mathrm{cyc},2}$. For $0.5 \leq M_{a}/M_{c} \leq 0.75$, in the two central panels of the bottom row in Fig. \ref{fig:accreted_mass}, the troughs merge. 
This coincides with the start of the nonlinear trend in magnetic dipole moment versus $M_{a}$; see Figure 11 in \citet{priymak2011}.
The separation in line energy $E_{\mathrm{cyc},2} - E_{\mathrm{cyc},1}$ decreases monotonically with $M_{a}$ for $M_{a} \ll M_{c}$, 
but the trend breaks down for $M_{a} \ga 0.75M_{c}$ for the same reason.

In reality, the temperature distribution at the surface of the accretion mound is not uniform;
the temperature is lower than it would otherwise be, wherever $|\mathbf{B}|$ is higher, because the magnetic field provides
pressure support, like in a Solar sunspot \citep{priest1982}.
A full calculation of thermal transport lies outside the scope of this paper.
To estimate the effect crudely, we multiply the flux emitted from each surface point by the Stefan-Boltzmann correction factor \citep{priest1982}
\begin{equation}
\lambda(\theta) = \left\{ 1 - \beta \mathbf{B}[r_{m}(\theta), \theta]^{2} \right\}^{4},
\label{magnetic_modulation} 
\end{equation}
where $\beta \propto (\rho T_{0})^{-1}$ depends on the temperature $T_{0}$ at the pole.
The second term in braces equals the ratio of the magnetic pressure at colatitude $\theta$ to the kinetic pressure at the pole,
where $|\mathbf{B}|$ is small. It is always positive for the typical magnetic fields and temperatures 
($10^{12} \ \mathrm{G} \la |\mathbf{B}| \la 10^{13} \ \mathrm{G}$, $T_{0} \sim 10^{8} \ \mathrm{K}$) relevant to CRSF formation in accreting neutron stars.
Equation (\ref{magnetic_modulation}) expresses the constancy of the total (kinetic plus magnetic) pressure in MHD equilibrium but does not
take into account corrections due to additional, ongoing, accretion-driven heating.

The results of applying the correction factor (\ref{magnetic_modulation}) to cyclotron spectra
are presented in Fig. \ref{fig:accreted_mass} as blue dashed curves.
In comparison to accretion mounds with uniform surface temperature, the `sunspot' effect suppresses the absorption feature at $E_{\mathrm{cyc},2}$ 
and augments the feature at $E_{\mathrm{cyc},1}$. 
The $E_{\mathrm{cyc},2}$ line originates near the magnetic equator, where $|\mathbf{B}|$ is enhanced by burial and $T$ is lower, 
whereas the $E_{\mathrm{cyc},1}$ line originates near the magnetic pole, where $T$ reaches its maximum value $T_{0}$ (see Fig. \ref{fig:surface}).

The above results suggest one qualitative way to discriminate between magnetic and thermal mountains in different classes of accreting neutron stars. 
The cyclotron spectrum of a magnetic mountain depends strongly on $M_{a}$, whereas the spectrum of a thermal mountain depends mainly
on accretion-driven heating, i.e. $\dot{M}$ and the hot-spot geometry \citep{ushomirsky2000}.
In young accretors, where insufficient mass has accreted to completely bury the magnetic field below the crust,
the CRSFs may evolve on the burial time-scale $M_{c}/\dot{M} \sim 10^{2} \text{--} 10^{3} \ \mathrm{yr}$.
Such evolution may be detectable over the next decade with next-generation X-ray telescopes.
In contrast, CRSFs from thermal mountains should remain more stable.
Needless to say, much more work (e.g. on thermal transport) is needed to establish the practicality of this idea.

\section{Equation of State}
\label{sec:equation_of_state}

\begin{figure*}
\begin{minipage}{175mm}
\subfigure
{
	\includegraphics[width=175mm]{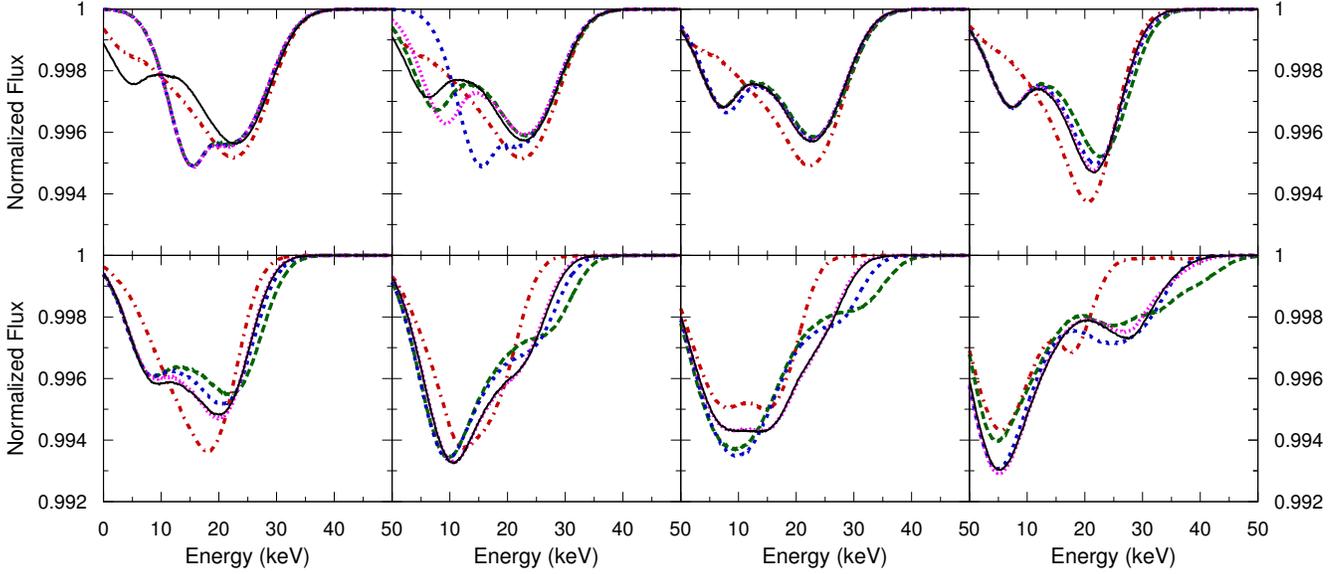}
}
\caption{CRSF spectra for a magnetic mountain in the aligned configuration $(\iota, \alpha) = (0,0)$ (i.e. no phase dependence), with $u=0.4$, 
$\psi_{\ast} = 10^{24} \ \mathrm{G} \ \mathrm{cm}^{2}$, and $M_{a}/M_{c} = 10^{-4}$, $10^{-3}$, $10^{-2}$, $10^{-1}$ 
(top row, left to right column), and $M_{a}/M_{c} = 0.25$, $0.5$, $0.75$, $1.0$ (bottom row, left to right column).
Each panel displays curves corresponding to EOS models A (dot-dashed red), B (short-dashed blue), 
C (dotted purple), D (long-dashed green), and E (solid black) \citep{priymak2011}.}
\label{fig:eos_spectra}
\end{minipage}
\end{figure*}

\begin{figure*}
\subfigure
{
	\includegraphics[width=84mm]{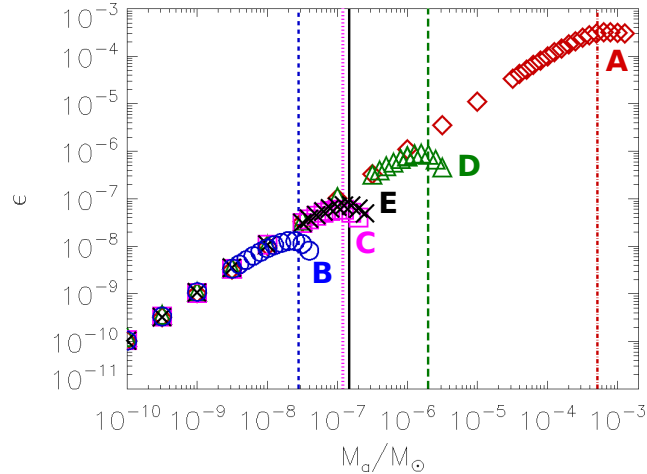}
}
\caption{Mass ellipticity $\epsilon$, as a function of accreted mass, $M_{\mathrm{a}}$,
measured in solar masses, for models A (red diamonds), B (blue circles), C
(purple squares), D (green triangles), and E (black crosses) [from \citet{priymak2011}]. Values of the characteristic mass
$M_{\mathrm{c}}$ for each model are drawn as vertical dot-dashed/short-dashed/dotted/long-dashed/solid lines and colour-coded accordingly.}
\label{fig:ellipticity_Ma}
\end{figure*}

In this section we examine how the 
EOS of the magnetized mound affects the cyclotron spectrum, to scope out the range of possibilities.
For the sake of definiteness, we analyse the polytropic EOS models B--D in \citet{priymak2011},
describing the various degenerate pressures in the neutron star crust, as well as an isothermal and mass-weighted EOS (models A and E respectively).  

In Figure \ref{fig:eos_spectra} we plot spectra from mounds for models A (dot-dashed red curve),
B (short-dashed blue curve), C (dotted purple curve), 
D (long-dashed green curve), and E (solid black curve)
as a function of $M_{a}$, with $M_{a}/M_{c} = 10^{-4}$, $10^{-3}$, $10^{-2}$, $10^{-1}$ 
(top row, left to right column), and $M_{a}/M_{c} = 0.25$, $0.5$, $0.75$, $1.0$ (bottom row, left to right column). 
The aligned configuration $(\iota, \alpha) = (0,0)$ is chosen deliberately to suppress the phase dependence, as in Fig. \ref{fig:accreted_mass}.

At each $M_{a}$, there are subtle differences in the spectra produced by different models.
The differences arise because the EOS affects the overall magnetic structure of the mound;
thermal EOS effects confined to the thin, X-ray-emitting skin of the mound are negligible.
Model A behaves qualitatively differently to models B--D.
For $M_{a} \ll M_{c}$, model A exhibits a single absorption trough, whereas models B--D exhibit two. 
As $M_{a}$ rises to $\approx 0.5 M_{c}$, the trough in model A becomes narrower, deeper, and shifts to lower energy.
On the other hand, the troughs in models B--D approach equal depths and subsequently merge.
For $M_{a} \ga 0.5 M_{c}$, model A splits into two distinct absorption features, whose energy separation increases with $M_{a}$,
with the lower energy feature becoming more prominent. In comparison, for models B--D, the CRSF splits
into two, and the lower energy feature dominates, as $M_{a}$ increases. 

The ellipticity $\epsilon$ is plotted as a function of accreted mass $M_{a}$ (in solar masses)
for models A--E in Fig. \ref{fig:ellipticity_Ma}. The results follow the approximate scaling 
$\epsilon = (M_{a}/M_{\odot})(1 + M_{a}/M_{c})^{-1}$ \citep{melatos2005, vigelius2009b}.
Since $M_{c}$ varies by a factor of $\sim 10^{4}$ between the models, so too does the value where $\epsilon$ saturates in Fig. \ref{fig:ellipticity_Ma}.
The $\epsilon(M_{a})$ relation in the figure, and the scaling of magnetic dipole moment $\mu$ with $M_{a}$ presented elsewhere \citep{payne2004,priymak2011},
combine to imply a relation $\mu(\epsilon)$ which is observationally testable in principle, if $\epsilon$ can be inferred 
from CRSF studies or a gravitational wave measurement (if the spin period and source distance are known).
By comparing observed cyclotron spectra with those predicted by models A--E as functions of $M_{a}$, we can, in
principle, extract information on the EOS of the accreted matter, stellar orientation, and surface magnetic field.
As in Section \ref{sec:accreted_mass}, however, there is a long way to go to establish the practicality of 
multimessenger investigations of this kind.

\section{Discriminating between thermal and magnetic mountains}
\label{sec:discriminating_mountains}

\begin{figure*}
\begin{minipage}{175mm}
\subfigure
{
	\includegraphics[width=84mm]{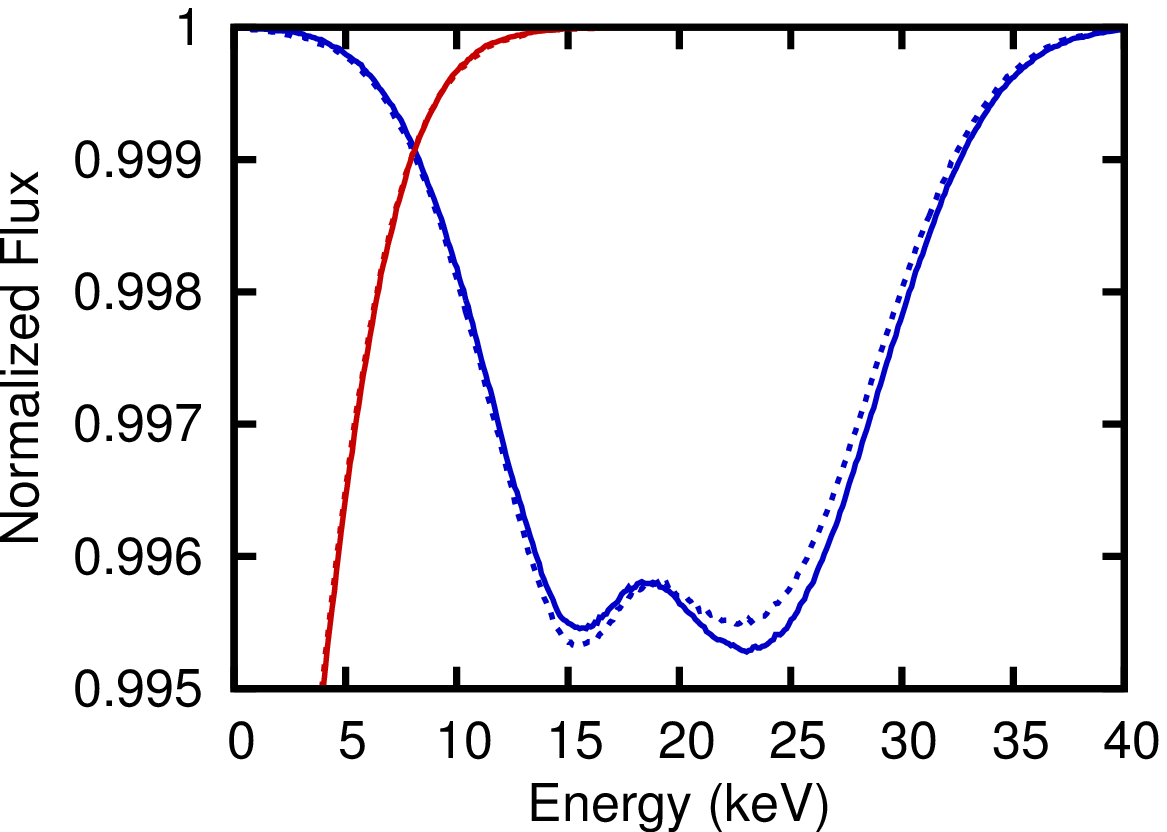}
}
\subfigure
{
	\includegraphics[width=84mm]{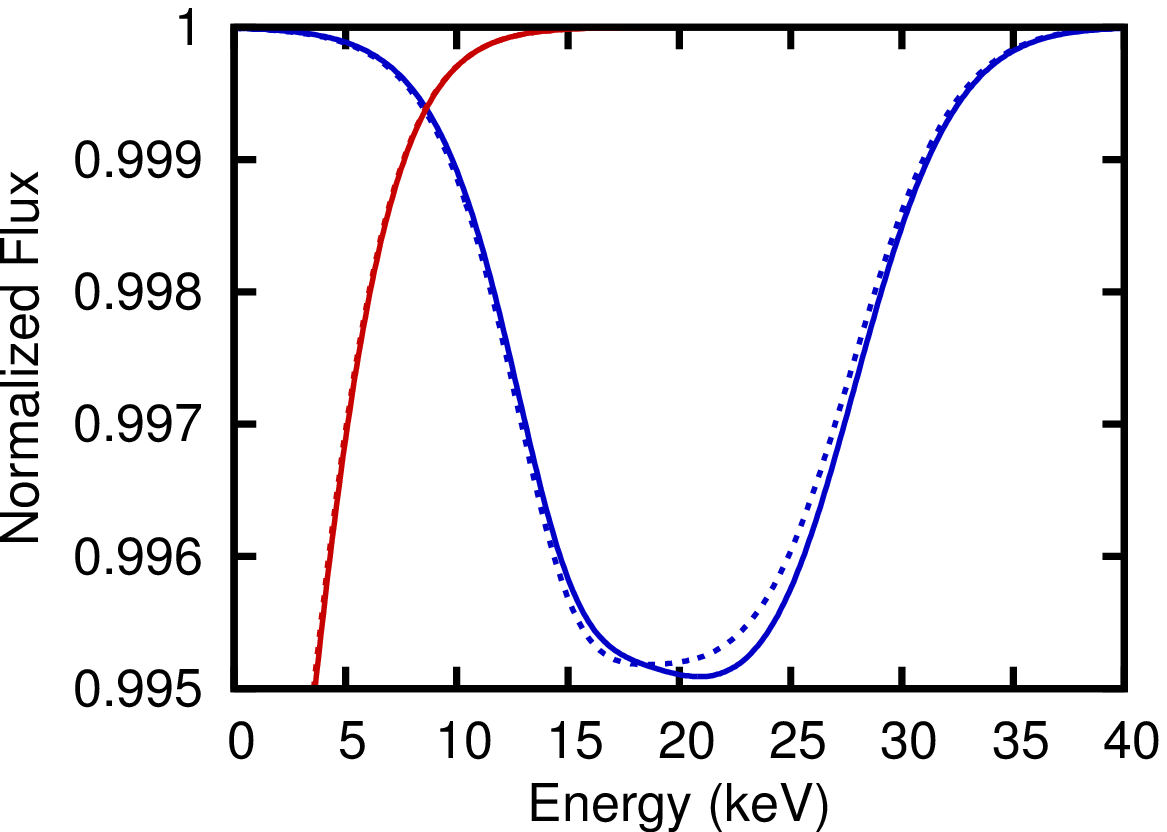}
}
\caption{Instantaneous ($\omega=3\pi/2$, left panel) and phase-averaged (right panel) cyclotron spectrum for 
scenario (\ref{a}) (thermal mountain without magnetic burial) with $(\iota, \alpha) = (\pi/4,\pi/4)$ and $u=0.4$. 
The results for strong and weak magnetic fields, and `hot spot' and quadrupole surface temperature distributions, are colour-coded as follows:
$\psi_{\ast} = 1.6 \times 10^{24} \ \mathrm{G} \ \mathrm{cm}^{2}$ (blue curves),
$\psi_{\ast} = 1.6 \times 10^{20} \ \mathrm{G} \ \mathrm{cm}^{2}$ (red curves), 
$(\delta T_{10}, \delta T_{20}) = (0.1, 0)$ (dotted curves), and 
$(\delta T_{10}, \delta T_{20}) = (0, 0.1)$ (solid curves).}
\label{fig:thermal_mountain_only}
\end{minipage}
\end{figure*}

\begin{figure*}
\begin{minipage}{175mm}
\subfigure
{
  \includegraphics[width=84mm]{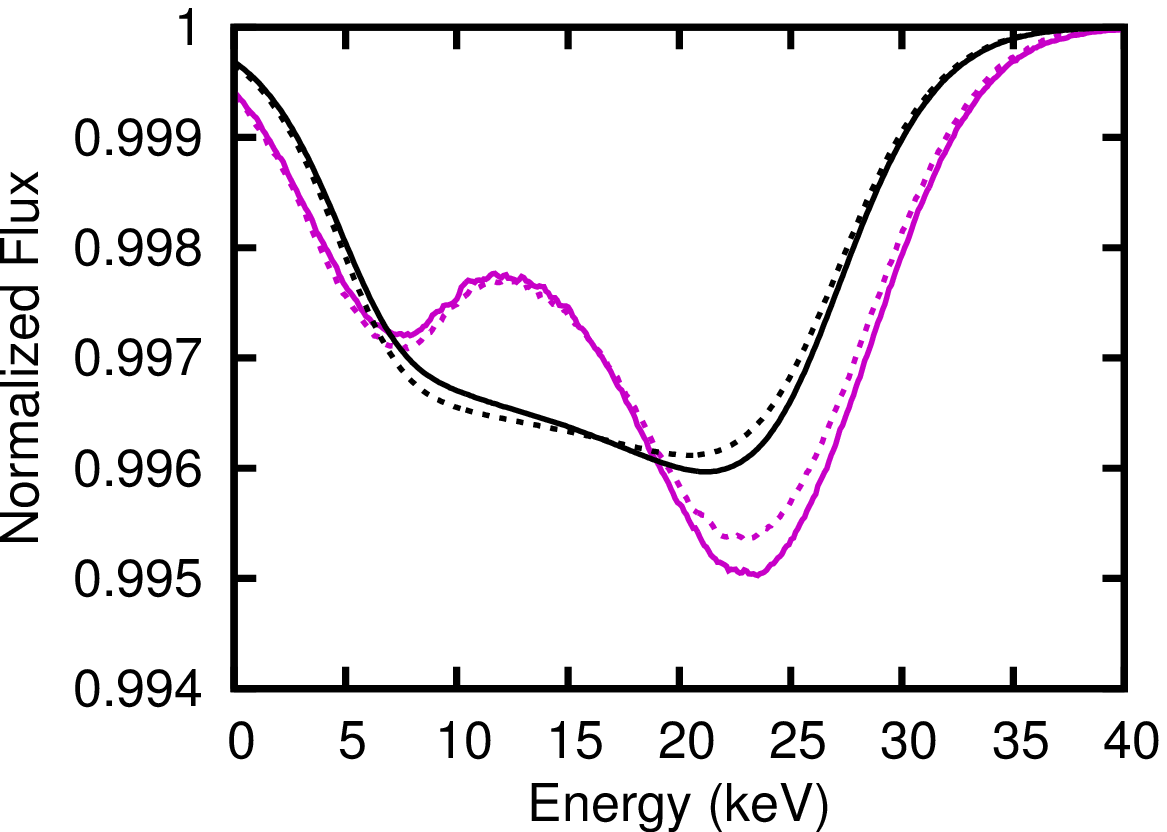}
}
\subfigure
{
  \includegraphics[width=84mm]{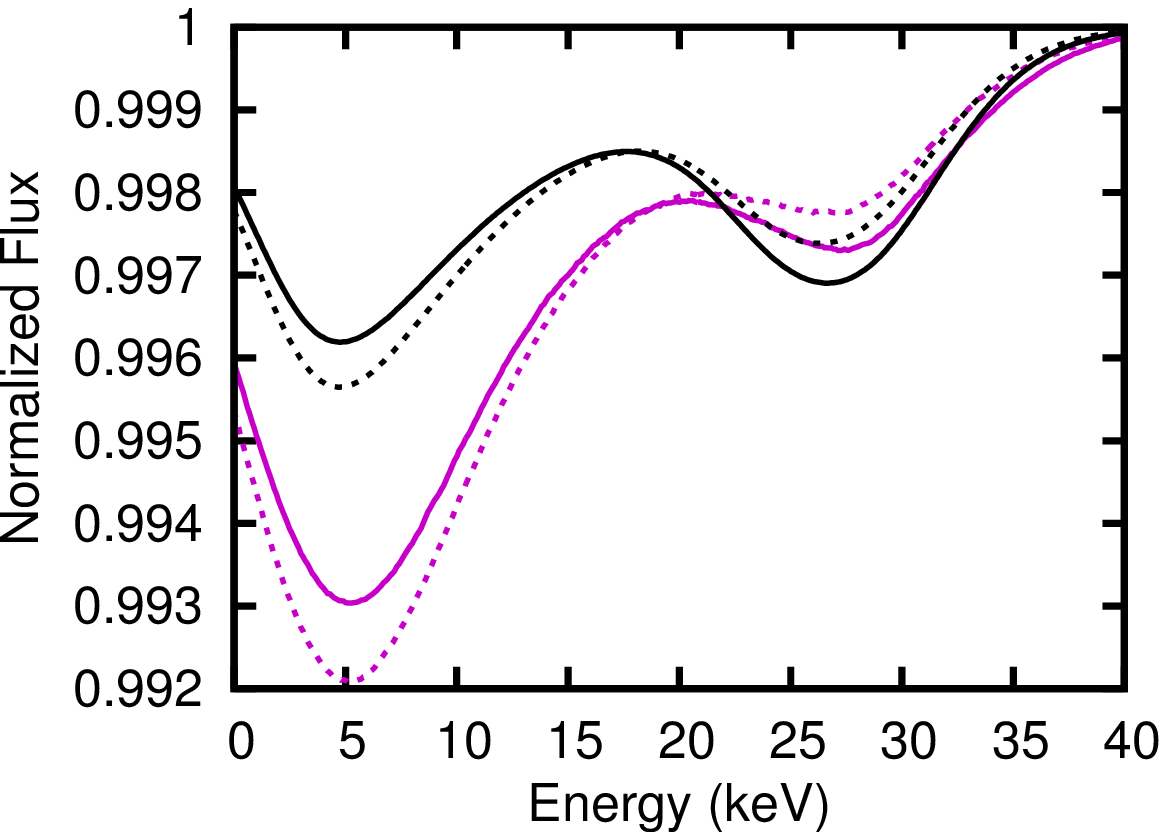}
}
\caption{
(\textit{Left panel}) Instantaneous ($\omega=3\pi/2$, purple curves) and 
phase-averaged (black curves) spectra for scenario (\ref{b}) (thermal mountain with magnetic burial) 
with $\psi_{\ast} = 1.6 \times 10^{24} \ \mathrm{G} \ \mathrm{cm}^{2}$, 
$(\delta T_{10}, \delta T_{20}) = (0.1, 0)$ (dotted curves), and $(\delta T_{10}, \delta T_{20}) = (0, 0.1)$ (solid curves). 
Other parameters as for Fig. \ref{fig:thermal_mountain_only}.
(\textit{Right panel}) Same as left panel, but for scenarios (\ref{c}) (solid curves) and 
(\ref{d}) (dotted curves) (magnetic mountain with uniform and magnetically modified surface temperature profile respectively).
}
\label{fig:magnetic_thermal_mountain}
\end{minipage}
\end{figure*}

\begin{figure*}
\begin{minipage}{175mm}
\subfigure
{
	\includegraphics[width=84mm]{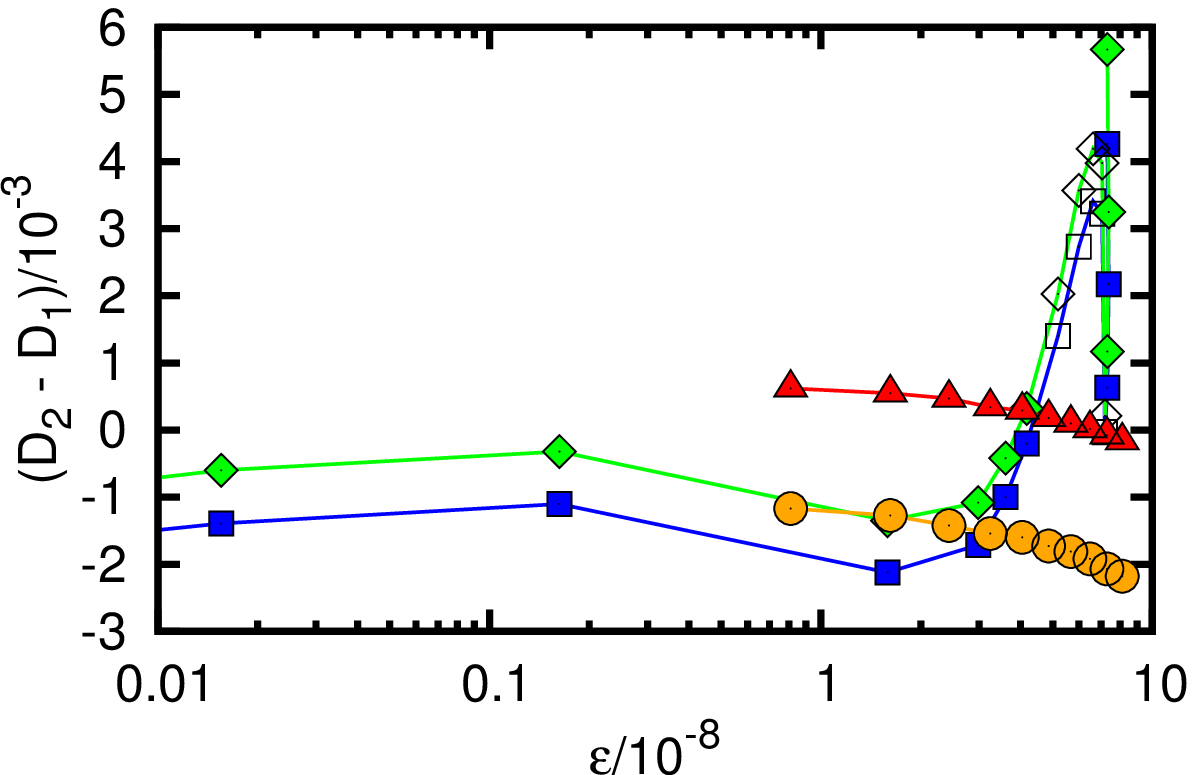}
}
\subfigure
{
	\includegraphics[width=84mm]{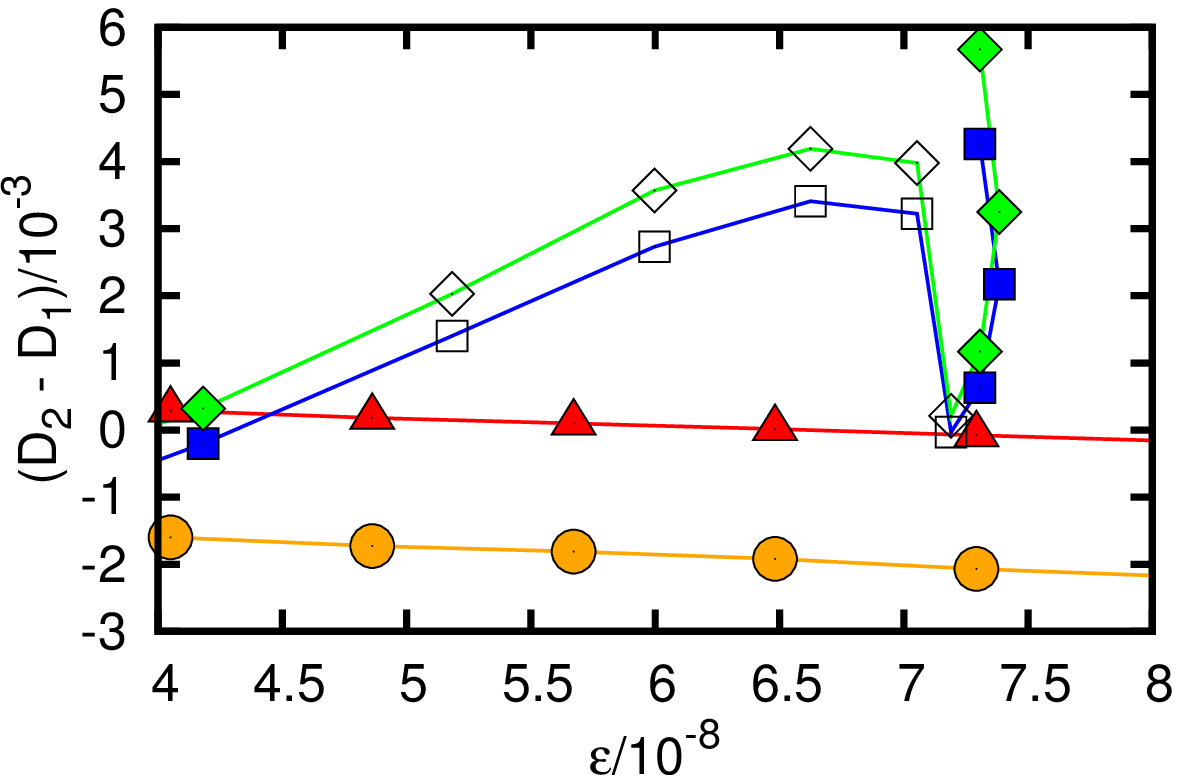}
}
\caption{Depth difference $D_{2} - D_{1}$ between the CRSF troughs at $E_{\mathrm{cyc},2}$ and $E_{\mathrm{cyc},1}$ (in normalized flux units) 
versus mass ellipticity for four different thermally and magnetically dominated mountain scenarios: (\ref{a}) with $\delta T = \delta T_{20}$ (red curve), 
(\ref{b}) with $\delta T = \delta T_{20}$ (orange curve), (\ref{c}) (blue curve), and (\ref{d}) (purple curve).
The right panel is the same as the left panel but zooms into the region $4 \leq \epsilon/10^{-8} \leq 8$ for clarity.
In the thermal mountain scenarios (\ref{a}) and (\ref{b}), $\epsilon(\delta T_{20})$ and $D_{2}(\delta T_{20}) - D_{1}(\delta T_{20})$ are computed 
for $0.01 \leq \delta T_{20} \leq 0.1$ (red triangles and orange circles respectively, left to right).
In the magnetic mountain scenarios (\ref{c}) and (\ref{d}), $\epsilon(M_{a})$ and $D_{2}(M_{a}) - D_{1}(M_{a})$ are computed for $10^{-3} \leq M_{a}/M_{c} \leq 1$
(blue squares and purple diamonds respectively, left to right).
Unfilled symbols indicate where the two CRSF troughs merge ($0.4 \la M_{a}/M_{c} \la 0.8$) and it becomes difficult to read off $D_{2} - D_{1}$.
In all cases, the instantaneous spectrum at $\omega=3\pi/2$ is used to illustrate the trends.}
\label{fig:thermal_magnetic_epsilon_DeltaD}
\end{minipage}
\end{figure*}

Mountains on accreting neutron stars are sustained either by magnetic stresses, as in Section \ref{sec:accretion_mound}, or by
thermoelastic stresses maintained by temperature gradients (e.g. from inhomogeneous heating in the crust),
as discussed in Section \ref{sec:introduction}.
Temperature gradients can also drive compositional gradients, e.g. `wavy' electron capture layers \citep{bildsten1998, ushomirsky2000}.
In this section we compare the signatures left by thermal and magnetic gradients on the cyclotron spectrum
and suggest observational methods for discriminating between them.
This is potentially interesting for studies of the surface magnetic structure of accreting neutron stars and also for 
objects that may be detected in future as gravitational wave sources.
The discussion is preliminary; a detailed exploration of the problem and its parameter space will be presented elsewhere (Haskell et al. 2014, in preparation).
We emphasize again that the CRSFs calculated below and elsewhere in this paper are too weak to be detected by currently operational X-ray telescopes.
The results in this section are a first attempt to think through what could be achieved by multimessenger observational strategies
in the context of thermal and magnetic mountains in the future.

\subsection{Scenarios}
\label{sec:scenarios}
Consider the four generic scenarios below for generating a mountain with ellipticity $\epsilon$ by accretion.

\begin{enumerate}[(a)]
\item Thermal gradients dominate $\epsilon$, and the process of magnetic burial described in Section \ref{sec:GS_equilibrium}
(mound spreads equatorward, $\mathbf{B}$ is compressed at the equator) does not occur. 
Upon decomposing the temperature
perturbation $\delta T = \sum_{l,m} \delta T_{lm}(r) Y_{lm}(\theta, \phi)$ into spherical harmonics $Y_{lm}$, with 
$T = T_{0} (1 + \delta T)$, we note that we can relate $\delta T$ to $\epsilon$ if $\delta T_{20}$ is the dominant term; see \citet{ushomirsky2000}.
$\delta T$ also modulates the surface X-ray flux and hence the instantaneous and phase-averaged cyclotron spectra. \label{a}

\item Thermal gradients dominate $\epsilon$, but magnetic burial does occur. Then the cyclotron spectrum (but not $\epsilon$) is modified
by magnetic burial. \label{b}

\item Magnetic stresses dominate $\epsilon$, and thermal gradients are negligible with respect to both $\epsilon$ and the surface X-ray flux;
i.e. we assume uniform surface temperature ($\delta T = 0$). Then the instantaneous and phase-averaged cyclotron spectra are affected by the distorted
magnetic field (i.e. $E_{\mathrm{cyc}}$, $F$, and $D$ depend on $M_{a}$; see Section \ref{sec:accreted_mass}).
As $\epsilon$ also depends on $M_{a}$, there are correlations between $E_{\mathrm{cyc}}$, $F$, $D$, and $\epsilon$. \label{c}

\item As for (\ref{c}), but the magnetic pressure modifies the temperature profile and hence the spectrum via equation (\ref{magnetic_modulation}),
without modifying $\epsilon$ significantly. \label{d}
\end{enumerate}

For thermally dominated mountains [scenarios (\ref{a}) and (\ref{b})], we set $\delta T_{lm} \leq 0.1$ \citep{ushomirsky2000},
and assume that $\delta T$ at the surface equals $\delta T$ in the crust (i.e. the heat flux is radial).
As the solutions in \citet{ushomirsky2000} do not depend on $m$, we specialize here to $m = 0$ for convenience, matching the 
assumption of axisymmetry in $\mathbf{B}$.

For the preliminary study below, we specialize to the orientation $(\iota, \alpha) = (\pi/4, \pi/4)$ for which the line parameters
of the phase-resolved spectra from magnetic and thermal mountains vary the most over one period (see Fig. \ref{fig:spectra(i,alpha)}),
as the LOS crosses the magnetic pole ($\hat{\mathbf{n}}_{o} \cdot \mathbf{M} = 1$) and equator ($\hat{\mathbf{n}}_{o} \cdot \mathbf{M} = 0$)
at $\omega=\pi/2$ and $\omega=3\pi/2$ respectively. The EOS in model E of \citet{priymak2011} is used throughout.

\subsection{CRSF signatures}
\label{sec:cyclotron_line_signatures} 
Let us begin by comparing the spectra for the `hot spot' ($\delta T \propto Y_{10}$) and quadrupole ($\delta T \propto Y_{20}$)
models in scenarios (\ref{a}) and (\ref{b}).
Fig. \ref{fig:thermal_mountain_only} displays the instantaneous (left panel, $\omega=3\pi/2$) and phase-averaged (right panel)
spectra emitted by a thermal mountain [scenario (\ref{a})] for 
$\psi_{\ast} = 1.6 \times 10^{24} \ \mathrm{G} \ \mathrm{cm}^{2}$ (black curves), $\psi_{\ast} = 1.6 \times 10^{20} \ \mathrm{G} \ \mathrm{cm}^{2}$ 
(red curves), $(\delta T_{10}, \delta T_{20}) = (0.1,0)$ (dotted curves), and $(\delta T_{10}, \delta T_{20}) = (0,0.1)$ (solid curves).
The left panel exhibits two troughs at $E_{\mathrm{cyc},1} \approx 15 \ \mathrm{keV}$ and $E_{\mathrm{cyc},2} \approx 23 \ \mathrm{keV}$.
The former is marginally deeper than the latter for $\delta T_{10} \neq 0$ and vice versa for $\delta T_{20} \neq 0$,
while the line widths are the same.
The right panel exhibits a single absorption feature at $\approx 18 \ \mathrm{keV}$ for $\delta T_{10} \neq 0$
and $\approx 23 \ \mathrm{keV}$ for $\delta T_{20} \neq 0$.
The pole dominates in the latter case.
In comparison, for $\delta T_{10} \neq 0$, the flux-dominating region shifts equatorward, where $|\mathbf{B}|$ is weaker.
The instantaneous spectrum exhibits two troughs in the narrow window $1.3 \la \omega/\pi \la 1.7$ and one trough 
elsewhere, so the single trough dominates the phase average.

At low magnetization, e.g. $\psi_{\ast} = 1.6 \times 10^{20} \ \mathrm{G} \ \mathrm{cm}^{2}$ (red curves in Fig. \ref{fig:thermal_mountain_only}), 
a single CRSF appears at $\la 5 \ \mathrm{keV}$ in both the instantaneous and phase-averaged spectra, with $D\approx0.99$ 
and $F\approx4 \ \mathrm{keV}$.
This feature is likely to be interpreted as the non-CRSF continuum in observational data.
Note that at low $\psi_{\ast}$ ($B/B_{\mathrm{crit}} \ll 10^{-2}$) our calculations lie outside the regime of validity of the scattering cross-section formulas 
in \citet{schonherr2007}.
A correct treatment falls outside the scope of this paper.

How does magnetic burial of the sort described in Section \ref{sec:accretion_mound} affect the CRSF from a predominantly thermal mountain?
The left panel of Fig. \ref{fig:magnetic_thermal_mountain} displays instantaneous ($\omega=3\pi/2$, purple curves) and 
phase-averaged (black curves) spectra for scenario (\ref{b}) with $\psi_{\ast} = 1.6 \times 10^{24} \ \mathrm{G} \ \mathrm{cm}^{2}$, 
$(\delta T_{10}, \delta T_{20}) = (0.1, 0)$ (dotted curves), and $(\delta T_{10}, \delta T_{20}) = (0, 0.1)$ (solid curves).
Compared to the right panel of Figure \ref{fig:thermal_mountain_only}, the single CRSF in the phase-averaged spectrum is wider 
($F\approx23 \ \mathrm{keV}$) and shallower ($D\approx0.996$), especially for $\delta T_{10} \neq 0$, to the point 
where the feature at $E_{\mathrm{cyc}} \approx 22 \ \mathrm{keV}$ is ill-defined.
In the instantaneous spectrum, magnetic burial increases the energy splitting 
($E_{\mathrm{cyc},1} \approx 8 \ \mathrm{keV}$, $E_{\mathrm{cyc},2} \approx 23 \ \mathrm{keV}$).
The $\delta T_{10}$ perturbation deepens and suppresses the feature at $E_{\mathrm{cyc},1}$ more than $E_{\mathrm{cyc},2}$,
because burial compresses the magnetic field 
(compare $E_{\mathrm{cyc},1}$ and $E_{\mathrm{cyc},2}$ in the left panels of Figs. \ref{fig:thermal_mountain_only} and \ref{fig:magnetic_thermal_mountain}).

In scenarios (\ref{c}) and (\ref{d}), magnetic burial dominates $\epsilon$ and also modifies the cyclotron spectrum. 
The surface temperature is uniform or magnetic-pressure-corrected respectively.
The right panel of Fig. \ref{fig:magnetic_thermal_mountain} displays the instantaneous ($\omega=3\pi/2$, purple curves) and 
phase-averaged (black curves) spectra for scenarios (\ref{c}) (solid curves) and 
(\ref{d}) (dotted curves). Both exhibit CRSFs at $E_{\mathrm{cyc},1} \approx 5 \ \mathrm{keV}$ 
and $E_{\mathrm{cyc},2} \approx 27 \ \mathrm{keV}$.
Magnetic pressure support suppresses $E_{\mathrm{cyc},2}$ and deepens $E_{\mathrm{cyc},1}$ (see Section \ref{sec:accreted_mass}),
but the magnitude of the effect is modest.

\subsection{Ellipticity}
\label{sec:ellipticity}
The ellipticity $\epsilon$ of an accretion mound can be related
to its CRSF properties through $M_{a}$, if the mountain is magnetically dominated, or $\delta T_{20}$, if it is 
thermally dominated.
Future gravitational wave detections of certain classes of accreting neutron stars potentially offer a way to probe this
relation and shed new light on accretion mound structure. 
In scenarios (\ref{a}) and (\ref{b}), for $\delta T_{20} \neq 0$, we can write \citep{ushomirsky2000}
\begin{eqnarray}
\label{ellipticity_T}
\epsilon & \approx & 8.1\times10^{-7} (R_{\ast}/10^{6} \ \mathrm{cm})^{2} (M_{\ast}/1.4 \ M_{\odot})^{-1} \nonumber \\
& & \times (T_{0}/10^{8} \ \mathrm{K}) (Q/50 \ \mathrm{MeV})^{3} \delta T_{20}, 
\end{eqnarray}
where $Q$ is the threshold energy for a single electron capture layer,  
neglecting sinking and shear.
In scenarios (\ref{c}) and (\ref{d}), we have $\epsilon \approx (M_{a}/M_{\odot})(1 - M_{a}/M_{c})^{-1}$ and $M_{c}$ given by equation (\ref{Mc}).
The properties of the CRSF also depend on $\delta T_{20}$ and $M_{a}$, as described in Section \ref{sec:cyclotron_line_signatures}.

Figure \ref{fig:thermal_magnetic_epsilon_DeltaD} plots the depth difference $D_{2} - D_{1}$ between the lines at $E_{\mathrm{cyc},2}$
and $E_{\mathrm{cyc},1}$ in the instantaneous ($\omega=3\pi/2$) spectrum for scenarios (\ref{a}) with $\delta T = \delta T_{20}$ (red curve), 
(\ref{b}) with $\delta T = \delta T_{20}$ (orange curve), (\ref{c}) (blue curve), and (\ref{d}) (purple curve).
In the thermal mountain scenarios (\ref{a}) and (\ref{b}), $\epsilon(\delta T_{20})$ and $D_{2}(\delta T_{20}) - D_{1}(\delta T_{20})$ are computed 
from equation (\ref{ellipticity_T})
for $0.01 \leq \delta T_{20} \leq 0.1$ (in $0.01$ increments; red triangles and orange circles respectively, left to right). 
In the magnetic mountain scenarios (\ref{c}) and (\ref{d}), $\epsilon(M_{a})$ and $D_{2}(M_{a}) - D_{1}(M_{a})$ are computed directly from 
the output of the GS solver (see Section \ref{sec:accretion_mound}) for $10^{-3} \leq M_{a}/M_{c} \leq 1$
(blue squares and purple diamonds respectively, left to right).
Unfilled symbols indicate where two CRSF troughs merge ($0.4 \la M_{a}/M_{c} \la 0.8$) and it becomes 
difficult to read off $D_{2} - D_{1}$; one of the minima effectively becomes a point of inflexion (see Section \ref{sec:accreted_mass}).

Even a small amound of accretion ($M_{a} = 10^{-2} M_{c}$) onto a thermal mountain buries the magnetic field sufficiently to decrease 
$D_{2} - D_{1}$ (orange curve and circles in Fig. \ref{fig:thermal_magnetic_epsilon_DeltaD}); cf. scenario (\ref{a}) (red curve and triangles).
Similarly, magnetically modifying the surface temperature through (\ref{magnetic_modulation}) produces an offset in $D_{2} - D_{1}$
compared to the case $\delta T=0$ in scenarios (\ref{c}) and (\ref{d}).
In scenarios (\ref{a}) and (\ref{b}), the trend in $D_{2} - D_{1}$ versus $\epsilon$ is monotonic and unaffected by magnetic burial, 
unlike in scenarios (\ref{c}) and (\ref{d}), where $D_{2} - D_{1}$ behaves nontrivially. 
Realistically, we expect magnetic and thermal mountains to co-exist in accreting neutron stars,
leading to a combination of scenarios (\ref{a})--(\ref{d}).
We postpone a more thorough analysis of these trends, including their dependence on orientation, to future work.

\subsection{Future observational targets}
\label{sec:future}
The accreting neutron stars which are expected to be the strongest emitters of gravitational radiation are LMXBs which possess spin periods 
$\ga 0.1 \ \mathrm{kHz}$ and weak dipole magnetic fields $\la 10^{9} \ \mathrm{G}$ assuming magnetocentrifugal equilibrium \citep{ghosh1979}. 
In contrast, objects which show evidence of CRSFs in existing data possess rotation periods of $10^{0} - 10^{2} \ \mathrm{s}$ and magnetic fields 
of $\sim 10^{12} \ \mathrm{G}$ (inferred from the CRSF line energy). Most CRSFs are in HMXBs with O- and B-type companions, but there are two notable 
exceptions --- Her X$-$1 and 4U 1626$-$67 --- which are LMXBs. 
It is commonly thought that the absence of cyclotron features in rapid rotators
is explained by their weak magnetization, but an alternative scenario is also possible (see Section \ref{sec:introduction}). A locally compressed, high-order-multipole magnetic field
(much stronger than the global dipole) can create a relatively large, gravitational-wave-emitting quadrupole ($\epsilon \la 10^{-7}$) while producing CRSFs which
are too weak to be detected by existing X-ray telescopes but may lie within reach of future instruments, as computed in Figures \ref{fig:thermal_mountain_only} and \ref{fig:magnetic_thermal_mountain}. 
This `magnetic fence' scenario (see also Section \ref{sec:introduction}) is supported by evidence that nuclear 
burning in weakly-magnetized systems is localized by compressed magnetic fields \citep{bhattacharya2006, misanovic2010, cavecchi2011, chakraborty2012}.
Combined GS and CRSF calculations indicate that CRSFs with line depths $\ga 0.1 \%$ can be obtained for $M_{a} \ \la M_{c}$ and dipole magnetic fields as low as $\sim 10^{11} \ \mathrm{G}$ \citep{haskell2014}.

We remind the reader that the CRSFs computed here are too weak to be detected by instruments currently in operation,
and that the computed ellipticities probably require next-generation gravitational wave detectors like the Einstein Telescope 
to be detected \citep{haskell2014}. 
Several planned X-ray missions are being designed to perform rotation-resolved spectroscopy of the thermal and non-thermal emission from neutron stars in the $0.1 \text{--} 10 \ \mathrm{keV}$ band.
The Neutron star Interior Composition Explorer (NICER) will target precise analysis of X-ray flux modulation due to rotating hot spots and absorption features near $\sim 1 \ \mathrm{keV}$,
achieving $\pm 5 \%$ uncertainties in derived quantities (eg. neutron star radius) for a $1 \ \mathrm{Ms}$ exposure at $\sim 10$ times the sensitivity of XMM-Newton \citep{gendreau2012}. This brings light curve 
modulations at the $\sim 0.3 \%$ level predicted by the theory (e.g. see Figure 6) into closer reach. Likewise, the Large Observatory for X-ray Timing (LOFT) is targeting $0.3 \ \mathrm{keV}$ spectral
resolution at $\sim 6 \ \mathrm{keV}$ and is planned to be sensitive to light curve modulations down to an amplitude of $\sim 0.4 \%$ for a $100 \ \mathrm{mCrab}$ source \citep{feroci2012}.

\section{Conclusion}
\label{sec:discussion}
In this paper we model the cyclotron spectrum emitted by a magnetically confined accretion mound on an accreting neutron star,
extending the pioneering calculations of \citet{mukherjee2012}.
We calculate the equilibrium hydromagnetic structure of the mound within the GS formalism, including 
an integral flux-freezing constraint.
We model the CRSF as a Gaussian absorption feature, incorporating the emission-angle dependence of CRSF properties,
gravitational light bending, and different surface temperature distributions characteristic of magnetically and thermally supported mounds.
We show that the instantaneous and phase-averaged spectra seen by an observer contain distinctive features, which can
be used in principle to discriminate between mountains of magnetic and thermal origin.

Polar magnetic burial influences the spectrum in the following important ways.

\begin{enumerate}
\item Orientation (Section \ref{sec:geometry}). 
If the observer and magnetic inclination angles satisfy $\pi/8 \la \iota, \alpha \la 3\pi/8$,
the phase-resolved spectrum transitions broadly between three states within one rotation period.
If either angle approaches $\pi/2$, the spectrum transitions between two distinct states.
If either angle approaches zero, the spectrum is phase invariant.
The constraints on $\iota$ and $\alpha$ thereby obtained are potentially useful
in gravitational wave applications.

\item Accreted mass (Section \ref{sec:accreted_mass}).
As $M_{a}$ increases, the lateral distortion and equatorial compression of the magnetic field modify
the spectrum strongly. For $M_{a} \ll M_{c}$, the CRSF properties vary with $M_{a}$
monotonically.
For $M_{a} \ga 0.5 M_{c}$, the trend breaks down; the troughs merge then separate again.
This trend is similar for uniform and magnetic-pressure-modified surface temperature distributions.
The low- and high-energy components emanate from the pole (low $|\mathbf{B}|$) and equator (high $|\mathbf{B}|$)
respectively. 

\item EOS (Section \ref{sec:equation_of_state}).
Two magnetic mountains with equal $M_{a}$ and different EOSs exhibit different CRSFs.
\end{enumerate}

CRSFs have been observed in $\sim 20$ accreting neutron stars, most but not all of which
are HMXBs \citep{pottschmidt2012, caballero2012}.
Generally but not exclusively, it is found that the profile of the fundamental line is asymmetric and not exactly Gaussian, 
the second harmonic is deeper than the fundamental, 
the line energy ratio is not harmonic,
and there is significant variation of CRSF properties with rotation phase and source luminosity.
Other authors have shown that the distorted magnetic field configuration arising from 
polar magnetic burial can explain some of these observations \citep{mukherjee2012}.
We confirm and extend that result here.
We also compare the CRSFs produced by thermal and magnetic mountains.
Phase-resolved spectra for both kinds of mountains display two troughs
at certain phases. After averaging over phase,
only one trough remains in a thermal mountain, whereas the magnetic mountain still
shows two (Section \ref{sec:discriminating_mountains}).
In systems expected to emit relatively strong gravitational wave signals,
perhaps detectable by future instruments like the Einstein Telescope,
we show that it is possible in principle to discriminate between mountains
of thermal and magnetic origin by correlating X-ray (e.g. line depth difference)
and gravitational wave (e.g. wave strain) observables in new multimessenger experiments.

The accretion mound and CRSF models discussed in this paper require further improvements before direct
quantitative contact with X-ray and gravitational-wave observations is possible. 
Time-dependent MHD simulations of thermal transport in mountains
are required to address the interaction between the distorted magnetic field
and accretion-induced heating, an important source of temperature asymmetries.
Extra physics like Hall drift (in the strong-$|\mathbf{B}|$ equator)
and Ohmic diffusion (in the polar hot spot) may enter, as well
as feedback between the accretion disk and surface mountain,
and directional radiative transfer in the accretion column (e.g. the `fan beam' in
subcritical accretors).  
Also, it is important to extend magnetic mountain simulations
to cover the full range of $M_{a}$ expected in accreting neutron stars (up to $\sim 10^{-1} M_{\odot}$ in LMXBs).
A magnetic mountain loses equilibrium for $M_{a} \geq 1.4 M_{c}$; there is no GS solution available and 
magnetic bubbles form, which pinch off above $M_{a} \approx 1.4M_{c}$ \citep{payne2004, mukherjee2012}.
This behaviour is not the same as a linear or resistive MHD instability (e.g. Parker), which
grows from an equilibrium state and possibly disrupts the mountain. Instead, there is
no equilibrium state at all \citep{klimchuk1989}, nor is the mountain disrupted;
the bubbles pinch off progressively as $M_{a}$ increases, keeping the mountain 
profile near its marginal structure at $M_{a} \approx 1.4 M_{c}$.
This process needs to be studied more. It mainly affects the lower-energy
CRSF trough, which is dominated by conditions at the magnetic pole.

We remind the reader that the CRSFs generated from thermal and magnetic mountains
are shallow and extremely challenging to observe with existing X-ray telescopes,
especially in the quiescent emission of accreting neutron stars with millisecond spin periods. 
This paper offers a proof of concept that multimessenger observations
with next-generation X-ray and gravitational wave observatories
can, in principle, elucidate the surface structure of accreting neutron stars and discriminate
between mass quadrupole moments of thermal and magnetic origin.

\section*{Acknowledgements}
The authors thank Brynmor Haskell for comments. 
This work was supported by a Discovery Project grant
from the Australian Research Council.

\bibliographystyle{mn2e}
\bibliography{bibliography}

\end{document}